\documentclass{aa}

\usepackage{graphicx}
\usepackage{amsmath}
\usepackage{subcaption}
\usepackage{mwe}
\usepackage{booktabs}
\usepackage{tabularx}
\usepackage{natbib}
\usepackage{xcolor}
\bibpunct{(}{)}{;}{a}{}{,}
\setcitestyle{citesep={;}}
\usepackage{gensymb}
\usepackage{graphicx}
\usepackage{txfonts}
\usepackage{siunitx}
\newcommand{\mockalph}[1]{}

\begin{document} 

   \title{The SRG/eROSITA All-Sky Survey: \\
 X-ray beacons at late cosmic dawn}
    \titlerunning{X-ray beacons at late cosmic dawn}
    \authorrunning{Wolf, Salvato, Belladitta et al.}

   \author{J. Wolf\,\thanks{jwolf@mpia.de}
          \inst{1,2,3},
 M. Salvato
          \inst{1,2},
 S. Belladitta
          \inst{3,4},
 R. Arcodia
          \inst{1,5}
 S. Ciroi
          \inst{6}, 
 F. Di Mille
          \inst{7}, 
 T. Sbarrato
          \inst{8},
 J. Buchner
          \inst{1} \\
 S. Hämmerich
          \inst{9},
 J. Wilms 
          \inst{9},
 W. Collmar
         \inst{1},
 T. Dwelly
          \inst{1},
 A. Merloni
          \inst{1},
 T. Urrutia
          \inst{10},
 K. Nandra
          \inst{1}
          }

   \institute{Max-Planck-Institut f\"{u}r extraterrestrische Physik, Gie\ss enbachstra\ss e 1, 85748 Garching, Germany
         \and
         Exzellenzcluster ORIGINS, Boltzmannstr. 2, 85748 Garching, Germany  
         \and
         Max-Planck-Institut f\"{u}r Astronomie, Königstuhl 17, 69117 Heidelberg, Germany 
         \and
         INAF — Osservatorio di Astrofisica e Scienza dello Spazio, via Gobetti 93/3, I-40129, Bologna, Italy
         \and
         MIT Kavli Institute for Astrophysics and Space Research, 70 Vassar Street, Cambridge, MA 02139, USA 
         \and Dipartimento di Fisica e Astronomia dell’ Universit\`{a} di Padova ,Vicolo dell’Osservatorio 3, I-35122 Padova, Italy
         \and Las Campanas Observatory - Carnegie Institution for Science, Colina el Pino, Casilla 601, La Serena, Chile
         \and INAF – Osservatorio Astronomico di Brera, Via Bianchi 46, 23807 Merate, Italy 
         \and Dr. Karl Remeis-Observatory, Friedrich-Alexander-Universit\"{a}t Erlangen-N\"{u}rnberg, Sternwartstr. 7, 96049 Bamberg,
Germany
         \and Leibniz-Institut für Astrophysik Potsdam (AIP). An der Sternwarte 16. 14482 Potsdam, Germany \\
             }

   \date{Received June 7, 2024; accepted }

  \abstract
   {The Spectrum Roentgen Gamma (SRG)/eROSITA All-Sky Survey (eRASS) is expected to contain $\sim100$ quasars that emitted their light when the universe was less than a billion years old, i.e. at $z>5.6$. By selection, these quasars populate the bright end of the AGN X-ray luminosity function, and their space density offers a powerful demographic diagnostic of the parent super-massive black hole population.  }
   {Of the $\gtrapprox 400$ quasars that have been discovered at $z>5.6$ to date, less than 15\% have been X-ray detected. We present a pilot survey to uncover the elusive X-ray luminous end of the distant quasar population.}
   {We have designed a quasar selection pipeline based on optical, infrared and X-ray imaging data from DES DR2, VHS DR5, CatWISE2020 and the eRASS (up to eRASS:4). 
   The core selection method relies on SED template fitting. We performed optical follow-up spectroscopy with the \textit{Magellan}/LDSS3 instrument for the redshift confirmation of a subset of candidates. We have further obtained a deeper X-ray image of one of our candidates with \textit{Chandra} ACIS-S.}
   {We report the discovery of five new quasars in the redshift range $5.6 < z < 6.1$. Two of these quasars are detected in eRASS and are, therefore, X-ray ultra-luminous by selection. We also report the detection of these quasars at radio frequencies. The first one is a broad absorption line quasar, which shows significant, order-of-magnitude X-ray dimming over 3.5 years, i.e. about six months in the quasar rest frame. The second X-ray detected quasar is a jetted source with compact morphology. We show that a blazar configuration is likely for this source, making it one of the most distant blazar known to date.}
   {With our pilot study, we demonstrate the power of eROSITA as a discovery machine for luminous quasars in the epoch of reionization. The X-ray emission of the two eROSITA detected quasars are likely to be driven by different high-energetic emission mechanisms, a diversity which we will further explore in a future systematic full-hemisphere survey.}


   \maketitle
%

\section{Introduction}

The study of accreting super-massive black holes (SMBHs) in the first gigayear after the Big Bang reshapes our understanding of black hole formation and growth, with potentially profound implications on the evolution of their host galaxies \citep[e.g.][]{habouzit22}. With the advent of deep infrared (IR) space-based imaging, as well as multi-object and slitless spectroscopy with the \textit{James Webb} Space Telescope (JWST), the redshift frontier of accreting SMBHs has been pushed far into cosmic dawn, to $z>8$ \citep[e.g.][]{harikane23,larson23,buncker23,maiolino23}. Combining low-resolution prism spectroscopy with JWST/NIRSpec and with the high-spatial-resolution of the \textit{Chandra} X-ray space telescope, \citet{goulding23} have recently discovered a (lensed) SMBH-powered active galactic nucleus (AGN) at z=10.1, corresponding to a lookback-time of $\sim$13.2 Gyr. 
The most extreme SMBHs arguably reside in the cores of quasars, luminous persistent panchromatic beacons, that are radiating up to $L_{\mathrm{bol}} \sim 10^{48} \mathrm{erg \, s^{-1}}$ at cosmic times reaching back deep into the epoch of reionization. While the less massive SMBHs discovered by JWST may grant us access to crisp snapshots of earlier evolutionary stages of AGN and their hosts, the mere existence of matured quasars and their central $\sim 10^9 \, M_\odot$ SMBHs in the first gigayear of the universe \citep[up to $z\sim 7.5$, e.g.][]{banados18,wang21,yang21} poses a puzzling progenitor problem: How did the seeds of black holes evolve to such massive systems in the cosmic timespan of a few hundred million years? Even under the unlikely assumption of sustained Eddington-limited accretion, well-understood stellar seeding models based on the collapse of early Pop III stars struggle to explain the formation of the most massive SMBHs found in the cores of $z>6.5$ quasars \citep[for reviews see][]{inayoshi20,fan23}. Due to the steep decline of the AGN population with luminosity at high redshift, quasars with  $L_{\mathrm{bol}} \sim 10^{48} \, \mathrm{erg \, s^{-1}}$ are rare in the first gigayear \citep{jiang16,vito18,kulkarni19,matsuoka23lum,schindler23}. The low space-density ($\sim 1 \; \mathrm{Gpc^{-3}}$) of these objects implies that wide-field surveys, sampling large cosmological volumes, are needed for their search. The optical and near-IR (NIR) \textit{Euclid} Wide Survey will enable the discovery of numerous $z>7.5$ quasars \citep{euclid19} in the coming years.

At slightly lower redshifts, the most luminous quasars at $z\sim 6$, impose less stringent constraints on black hole seed models but set a strict lower limit on the black hole population in the early universe and their accretion history \citep{vito18,wolf21,wolf22,barlow_hall22}. These rare AGN populating the extreme end of the luminosity and mass scale constitute powerful demographic probes \citep[e.g.][]{belladitta20}.
So far, about $\sim 475$ reionisation-era quasars have been discovered in wide optical and IR surveys \citep[][and references therein]{fan23}. Selection methods are based on the blue-band dropout photometry imprinted by neutral hydrogen absorption by the intergalactic medium \citep[IGM][]{madau00}. Quasar searches in imaging data are challenging as their optical and IR colours are similar to high-surface-density Galactic contaminants: late-type stars \citep[M-, L- and T-dwarfs, e.g.][]{caballero08}. 
M-dwarfs, the most numerous stars in the galaxy \citep{Chabrier01}, emit in the X-rays due to coronal activity and flares \citep[e.g.][]{briggs04,rutledge04,stelzer13,maggauda22}. However, at $z>5.7$, the bulk of galactic colour-contaminants has spectral type SpT M7 or later \citep{yang17}. These objects, dubbed ultra-cool dwarfs (UCDs), are intrinsically X-ray faint in their quiescent state. \citet{Stelzer22} have investigated the X-ray emission of UCDs in the first pass all-sky survey \citep[eRASS1\footnote{The number 1 refers to the number of all-sky surveys}][]{merloni24} of the extended R\"ontgen Survey with an Imaging Telescope Array \citep[eROSITA,][]{predehl21} onboard the Spektrum-R\"ontgen-Gamma (SRG) mission \citep{sunyaev21}. Out of 4754 spectroscopically confirmed UCDs (14914 candidates) in their sample, 21 (37) are associated with an eRASS1 source, i.e. less than $0.3 \%$ of the input catalogue. \citet{Stelzer22} show that the eRASS1-detected UCDs have large X-ray to bolometric luminosity ratios (log $L_\mathrm{X}/L_\mathrm{bol} \sim -3$), indicative of flaring activity. They conclude that the eRASS1 survey is only sensitive to flaring or the closest quiescent UCDs ($<20 \, \mathrm{pc}$). We note that a volume complete X-ray survey of nearby M-dwarfs ($<10 \, \mathrm{pc}$) based on eRASS:4 was presented by \citet{caramazza23}.

Thus, the detection of X-rays in eROSITA can significantly reduce the degeneracy between UCDs and high-redshift quasars using the latter's ubiquitous, strong coronal and jet X-ray emission. In this work, we present a pilot $z\sim 6$ quasar discovery program in eRASS1 through eRASS:4. \citet{wolf21} have already demonstrated the sensitivity of eROSITA to the X-ray emission of $z\sim 6$ quasars: we expect eROSITA to detect $\sim 100$ of the most X-ray luminous quasars ($\mathrm{log} \, L_\mathrm{X} > 45.5$), at the end of the mission (eRASS:8). In this luminosity-redshift regime, the strong X-ray emission arises from a range of high-energetic physical effects including Doppler boosting of non-thermal jet emission \citep[e.g.][]{belladitta20}, inverse Comptonization of cosmic microwave background photons \citep[iC-CMB, e.g.][]{ighina19,medvedev20} and rapid accretion \citep[e.g.][]{wolf22,zappacosta23}. eROSITA thus probes a diversity of X-ray boosting mechanisms whose contributions are yet to be disentangled \citep[e.g.][]{ighina21}. Our selection pipeline combines optical, IR and X-ray information. Similar panchromatic searches, spanning multiple orders of magnitudes in frequency, have recently been successfully performed by \citet{gloudemans22,ighina23}. In these works, the authors based their selection on the detection of jets at radio frequencies with the LOw-Frequency ARray (LOFAR, \citealt{vanhaarlem13}) and the  Australian square kilometre array pathfinder (ASKAP, \citealt{hotan21}) instead of accretion signatures in X-rays. We report the discovery of 5 new $z>5.6$ quasars in the joint Dark Energy Survey data release 2 (DES DR2\footnote{In this work, DES always refers to the second data release of the survey.}, \citealt{abbott16}) footprint, two of which are robustly detected in the eRASS catalogues. The newly discovered eRASS quasars are rare and highly energetic phenomenons at high redshift, which we showcase in this work: an X-ray variable broad absorption line (BAL) quasar and a blazar, i.e. a quasar for which the jet axis closely aligns with the line of sight.

This paper is organised as follows. In Section \ref{sec:two} we present the optical, IR, and X-ray based quasar selection pipeline. Discoveries through spectroscopic confirmation are presented in Section \ref{sec:three}. In Section \ref{sec:five}, we present the properties of the two newly discovered eRASS quasars. We discuss the implication of these discoveries and present our conclusions in Section \ref{sec:six}. Throughout this work, magnitudes are given in the AB reference system. Unless specified otherwise, uncertainties are given at the $1\sigma$ level. Distance calculations assume a flat $\Lambda$CDM cosmology \citep[][$H_0=67.66 \pm 0.42 \, \mathrm{km/(Mpc \,s)}$, $\Omega_{m}=0.3111 \pm 0.0056$, $\Omega_{\Lambda}=0.6889 \pm 0.0056$, $\Omega_{K}=0.001 \pm 0.002$]{planck18} 
\section{Panchromatic selection pipeline}
\label{sec:two}
\subsection{Candidate selection in optical/IR}
\label{sec:oir}
We have selected $z>5.6$ quasar candidates using optical data from DES. In addition, NIR and mid-IR (MIR) photometry from the VISTA Hemisphere Survey \citep[VHS,][]{Mcmahon13} Data Release 5 and CatWISE2020 \citep{Marocco21} complement the DES data in our selection pipeline. The combination of optical and infrared broad-band photometry covers a significant portion of the $z>5.6$ quasar restframe UV/optical SED, and its efficiency for quasar discovery has previously been demonstrated \citep[e.g.][]{reed19}.

\subsubsection{Colour pre-selection}
\label{sec:colour_presel}
The main DES source catalogue contains close to 700 million sources. To increase the computational efficiency of our selection pipeline, we have devised a set of loose colour and magnitude cuts to reduce the initial DES sample to photometric \textit{i}-band dropouts. We impose the following cuts on the \textit{grizY} Kron photometry (\texttt{mag\_auto}) of the main DES source catalogue:

\begin{align*}
    &(\mathtt{mag\_auto\_i}-\mathtt{mag\_auto\_z} > 0.8  \\
    & \& \,\mathtt{mag\_auto\_z}-\mathtt{mag\_auto\_y} < 0.12 \\
    & \& \,\mathtt{mag\_auto\_r}> 22.5) \\
    & \mathtt{OR}  \\
    & (\mathtt{mag\_auto\_i}-\mathtt{mag\_auto\_z} > 2.2  \\
     & \&  \mathtt{magerr\_auto\_r}> 0.36)
\end{align*}

\begin{figure}
    \centering
    \includegraphics[scale=0.34]{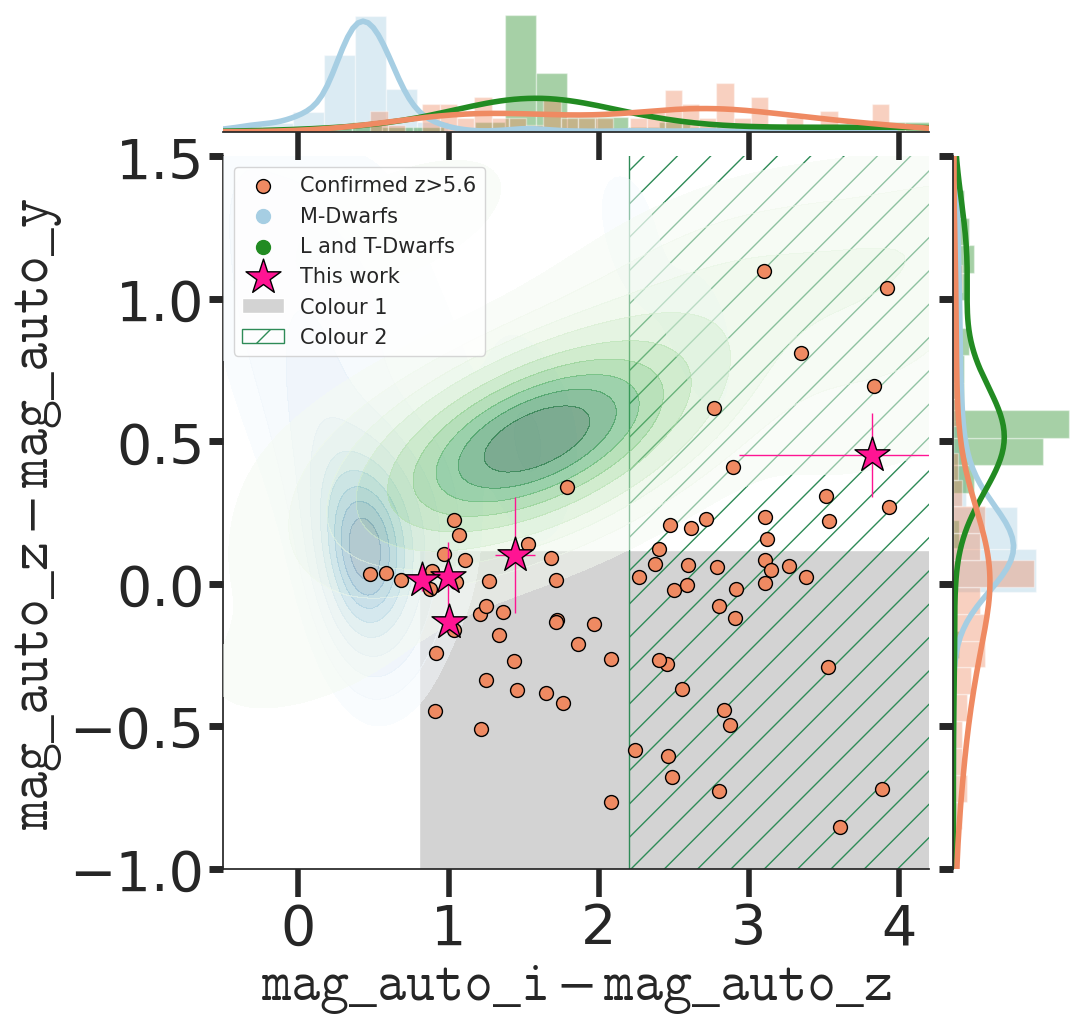}
    \caption{Colour-selection diagram. The hashed and grey regions correspond to the colour space from which we pre-selected high-redshift quasar candidates for the present pilot survey. Gaussian Kernel density estimates of DES detected M-dwarfs \citep{best18} and L/T-dwarfs \citep{liu02,hawley02,cruz03,cruz07,chiu08,reid08,schmidt2010,burningham10,kirkpatrick10,kirkpatrick11,wahhaj11,burningham13,dayjones13,marocco13,marocco14,marocco15,best15,cardoso15,tinney18} are displayed as shaded green and shaded blue areas, respectively. DES detected, spectroscopically confirmed $z>5.6$ quasars compiled in \citet{fan23} and additional sources from \citet{yang24} are shown as salmon red circles. 1D colour distributions of the samples from the literature are shown in the marginal histograms. Magenta stars indicate new quasars discovered in this work. Four of these discoveries are located at the edge of the grey selection box, close to the contamination region of brown dwarfs.}
    \label{fig:cc_diagram}
\end{figure}

In addition, we always impose signal-to-noise thresholds on the \textit{z}-band photometry (S/N > 10) and on the \textit{Y}-band photometry (S/N > 5). We further require no detection in the \textit{g} band (S/N < 3). These colour cuts are summarized in Fig. \ref{fig:cc_diagram}.
The colour cuts are more relaxed than the ones used in previous $z\sim 6$ quasar searches based on colour criteria \citep[e.g.][]{fan01,mortlock11,reed15,reed19,banados22,matsuoka22}, as the aim here is merely to pre-select dropout objects. We query a sub-sample of 511,654 candidates. 

\subsubsection{SED template fitting}

As the main candidate filtering step, we opted for spectral energy distribution (SED) template fitting and photometric redshift computation. 
Again the DES magnitudes in the bands \textit{g},\textit{r} ,\textit{i},\textit{z} and \textit{Y} are derived using the Kron aperture radius (\texttt{mag\_auto}).
In addition to the DES  photometry, we obtained magnitudes in the NIR and MIR wavebands by astrometrically cross-matching the main DES catalogue with the VHS and CatWISE2020 source catalogues within $<1''$. By construction, we only consider sources simultaneously detected in DES, VHS and CatWISE2020, reducing our sample to 220,221 sources. Combining the relatively deep DES optical photometry with these shallower IR surveys biases our selection towards very red, potentially dusty sources. Using the photometric code \texttt{Le PHARE} \citep[v2.2,][]{Arnouts11}, we performed a $\chi^2$-based template fitting analysis on the collected photometry of the colour pre-selected sample (see Section \ref{sec:colour_presel}). All magnitudes are directly read from the catalogues. For the VHS bands \textit{Y}, \textit{J}, \textit{H} and \textit{Ks} we have used the default point source aperture corrected magnitudes (aperture 3, 2’’ in diameter). Finally, for CatWISE2020 MIR magnitude measurements, we have selected profile-fit photometry corrected for proper motion in the bands W1 and W2 (\texttt{w1mpropm} and \texttt{w2mpropm}).

\texttt{Le PHARE} was supplemented with a custom template library for AGN and galaxies already used for the eROSITA Final Equatorial-Depth Survey by \citet{salvato21}. The original AGN templates used in this work are from \citet{polleta07,salvato09,brown19}. We also used mixed AGN-galaxy templates from \citet{salvato09} and \citet{ananna17}, based on inactive galaxy templates from \citet{noll04}. Due to their steep red spectra, UCDs of the spectral classes M, L and T are known to be the most prominent contaminants in photometric searches of neutral-IGM absorbed, high-redshift quasars, as both classes of objects appear as dropouts in optical bands. The spectral slope in the IR bands can help disentangle the typical quasar power-law accretion disk emission from the Rayleigh-Jeans tails of brown dwarf spectra \citep[for a review and quasar searches and contamination issues see][]{fan23}. In addition to stellar templates from \citet{Pickles98} and \citet{Bohlin95}, we account for the spectral diversity of these Galactic colour-contaminants by including the stellar templates provided by \citet{Chabrier00,knapp04,golimowski04,chiu06,bayo11}. 

\begin{figure}
    \includegraphics[scale=0.24]{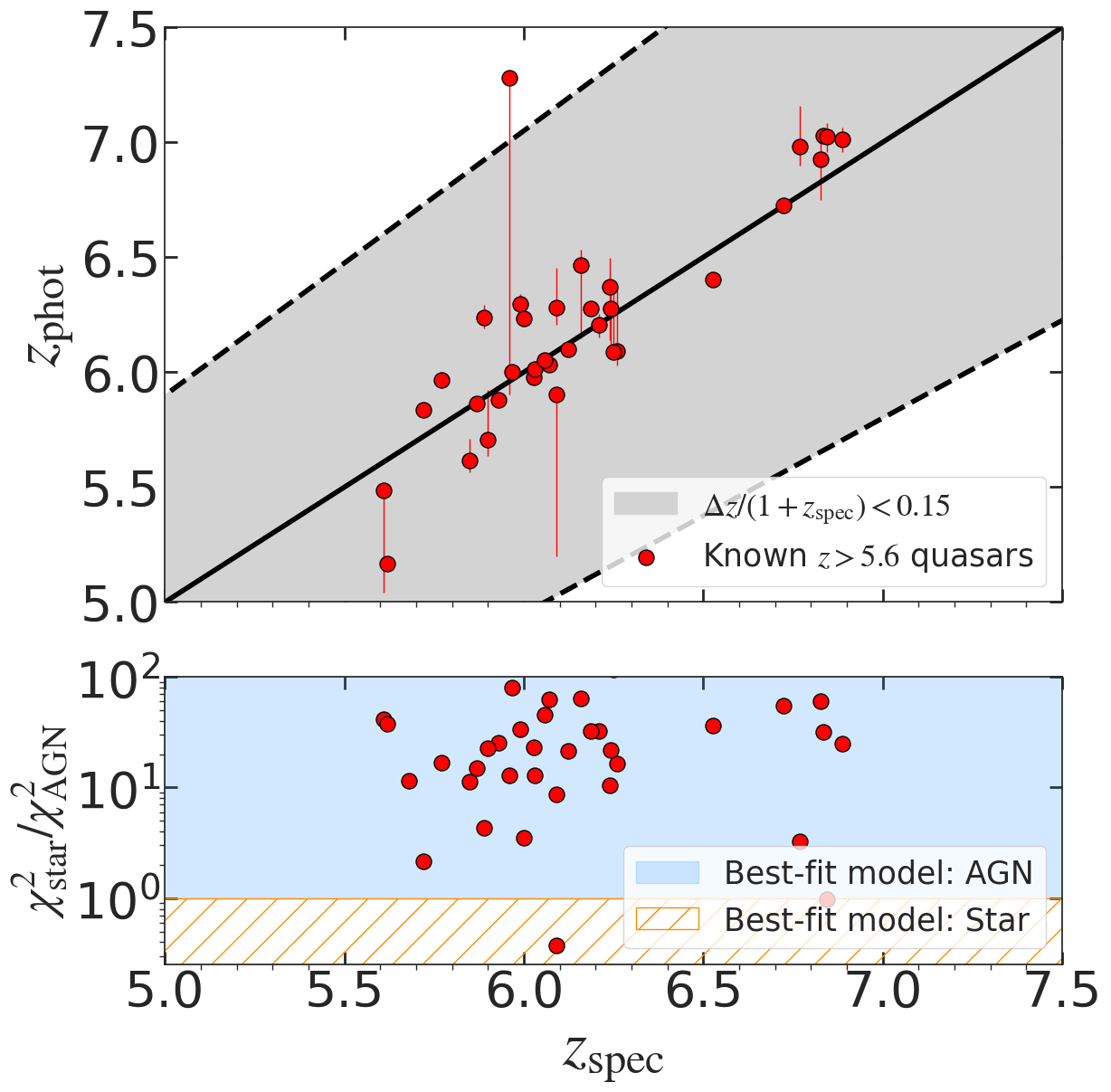}
    \caption{\texttt{Le PHARE} test run on previously spectroscopically confirmed $z>5.6$ quasars. \textit{Upper panel:}} The estimated photometric redshift $z_{\rm phot}$ is compared to the true spectroscopic value $z_{\rm spec}$. The grey shaded area indicates the region within which $\Delta z <0.15\times (1+z_\mathrm{spec})$ where $\Delta z = \mid z_\mathrm{spec} - z_\mathrm{phot}\mid$. \textit{Lower panel:} The ratio of $\chi^2$ of the stellar and the AGN or galaxy template are shown in the lower panel. In only two cases, the stellar solution is preferred.
    \label{fig:lephare_test}
\end{figure}

We tested our SED fitting analysis on all spectroscopically confirmed quasars at $z>5.6$ listed by \citet{fan23}, with detections within $<1''$ in DES, VHS and CatWISE2020. This sample contains 35 quasars with spectroscopic redshifts $5.61<z<6.89$ \citep[initial discoveries by][]{mortlock09,banados16,reed17,yang17,chehade18,decarli18,reed19,shen19, eilers20,venemans20,yang21,matsuoka22,banados22}. The \texttt{Le PHARE} photo-z calculating sub-routine \texttt{zphota} returns the $\chi^2$ values for the best-fit AGN and stellar templates. Out of the 35 currently confirmed DES, VHS and CatWISE2020 $z>5.6$ quasars, 34 are assigned a photometric redshift that follows the outlier rejection criterion $\mid z_\mathrm{spec} - z_\mathrm{phot} \mid /(1 + z_\mathrm{spec})< 0.15$ within its 1$\sigma$ uncertainties.  \citep[e.g.][]{hildebrandt01,salvato21} of which 32 have been fitted with AGN models with reduced $\chi_{\rm best}^2$ smaller than the reduced $\chi_{\rm star}^2$ values of the stellar models, i.e. the AGN model is favoured. The comparison of our photometric estimates of the redshift of these quasars and their actual spectroscopic redshifts is presented in Fig. \ref{fig:lephare_test}. 

We also ran the SED fitting routine on the  DES, VHS and CatWISE2020 photometry of known UCDs  \citep{liu02,hawley02,cruz03,cruz07,chiu08,reid08,schmidt2010,burningham10,kirkpatrick10,kirkpatrick11,wahhaj11,burningham13,dayjones13,marocco13,marocco14,marocco15,best15,cardoso15,tinney18}. Out of 51 late-type M, L and T dwarfs, 47 are correctly classified as stellar objects by \texttt{Le PHARE}.

\begin{figure}
    \includegraphics[scale=0.3]{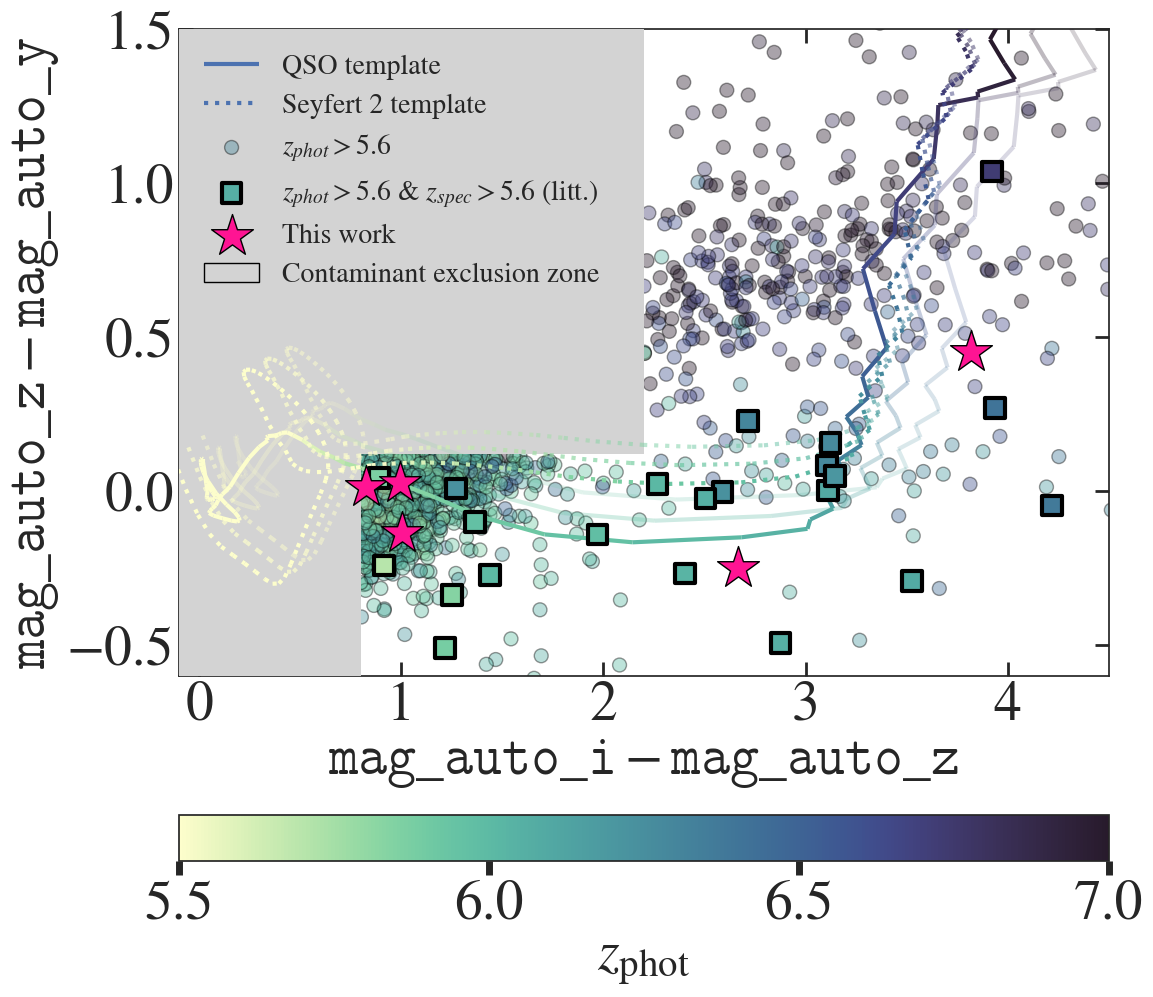}
    \caption{Optical colour-plane of selected of 2605 $z_\mathrm{phot}>5.6$ quasar candidates from our colour-cuts and SED fitting. The candidates (circles) are colour-coded according to their photometric redshift. We show redshift-colour tracks for \texttt{Le PHARE} templates: a typical quasar (pl\_QSO\_DR2\_029\_t0, an SDSS-DR5 composite modified by \citealt[]{salvato09}) and a Seyfert II template (Sey2\_template\_norm, \citealt{polleta07}), with increasing extinction values (0.0, 0.1, 0.2) displayed by increasing transparency. Squares indicate selected $z_\mathrm{phot}>5.6$ candidates that are already confirmed in literature \citep[references in][]{fan23}. The five quasars discovered in this work are shown as magenta stars.}
    \label{fig:sed_result}
\end{figure}

Computing photometric redshifts on the colour pre-selected sample of 220,221 sources with matches in VHS and CatWISE2020, we obtain a list of 2605 $z_\mathrm{phot}>5.6$ candidates. Their colour-redshift distribution is shown in Fig. \ref{fig:sed_result}. 

\subsubsection{Sample cleaning}

We have matched our quasar candidate sample of 2605 sources to compilations of known high-redshift quasars and UCDs. We had blindly selected 27 previously confirmed quasars at $5.7<z<6.8$ from \citet{mortlock09,jiang08,banados16,reed17,decarli18,chehade18,reed19,eilers20,venemans20,matsuoka22} and \citet{banados22}, i.e. 77 \% of all currently confirmed quasars in the joint DES, VHS and CatWISE2020 surveys. Additionally, our sample contained 7 known M, L and T dwarfs from \citet{liu02,tinney18} and \citet{best18}. These sources are discarded. We also removed one known low-redshift galaxy from \citet{colless01} from our sample.

We further extracted forced photometry on optical DES images of the quasar candidates to identify artefacts and problematic blue-band PSF-matched photometry \citep[e.g.][]{banados16,banados22}. Circular apertures of radius 3'' were centred on the optical coordinates of the candidates. We discard candidates for which we measure a significant aperture flux in g-band, i.e. with errors on g-band aperture magnitudes $\mathtt{ap\_magerr\_g}>0.36$ when a positive g-band flux is measured. Given the size of the source extraction region compared to the size of the DES point-spread-function (PSF $<1.11''$, \citealt{Abbott21}), this step effectively removes photometric dropouts that lie close to (or in close pairs with) the g-band detected stars and lower-redshift galaxies. Example images of discarded candidates are presented in Appendix \ref{append:images}. Following this step and a careful visual inspection of each candidate, we conserve a sample of 2098 quasar candidates.

\subsection{eROSITA down-selection}
\label{sec:erosita}
\begin{figure}
    \includegraphics[scale=0.50]{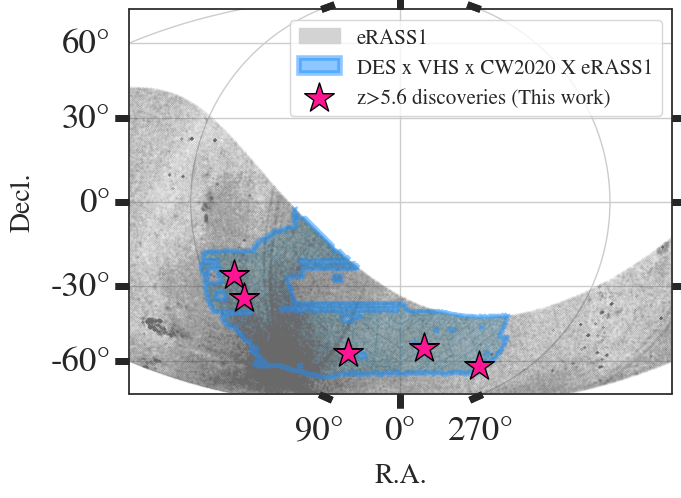}
    \caption{The intersection of the DES, VHS, CatWISE2020 and eRASS footprints (blue shaded area) covers a sky surface of $\sim 3500 \, \mathrm{deg^2}$. The underlying eRASS1 source population is shown in grey. The quasars discovered in this work are marked as pink stars.}
    \label{fig:footprint}
\end{figure}

We search for X-ray luminous quasars in eRASS \citep{merloni24}, and we therefore only account for quasar candidates contained in the intersection region of the optical/IR and eRASS footprints (1604 candidates). The combined area of our survey is measured with a coarse multi-order coverage using the \texttt{python} package \texttt{pymoc}. It spans $3507 \, \mathrm{deg^2}$. The survey footprint and the on-sky distribution of quasars discovered in this work are shown in Fig. \ref{fig:footprint}.

We verified whether our candidates were X-ray detected in any of the eRASS scans (eRASS1-4) or the deepest, complete cumulative survey eRASS:4. The eROSITA data analyzed in this work were processed with the standard processing pipeline \citep[][processing version 020]{brunner21}.
We performed forced photometry on the eRASS1-4 and eRASS:4 maps at the coordinates of the quasar candidates following a method presented by \citet{arcodia24}. In each eRASS scan and the deepest available cumulative image (up to eRASS:4), we extracted counts within a circular source aperture of radius 30'', corresponding the half-energy width (HEW) of the eROSITA PSF in the most sensitive energy range \citep[0.2-2.3 keV][]{dennerl12,predehl21,brunner21,merloni24}. We note, however, that the HEW of the PSF increases to 34.4'' in the 2.3-5.0 keV range. Additionally, background photons are extracted in an annulus of standard radii 4 times and 12 times the source region radius. Significantly detected sources in the background region are masked. The binomial no-source probability $P_b$ \citep[e.g.][]{weisskopf07,vito19} is given by:

\begin{equation}
    P_b(i\geq s) =  \sum_{i=s}^{s+b} \frac{(s+b)!}{i!(i-s-b)!} \left( \frac{1}{1+r} \right)^i\left(\frac{2+r}{1+r}\right)^{s+b-i}  
    \label{eq:nosource}
\end{equation}

\begin{figure}
    \includegraphics[scale=0.48]{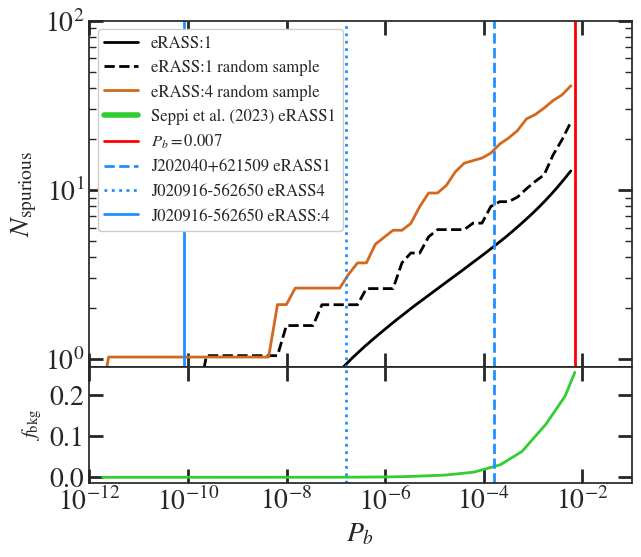}
    \caption{\textit{Upper panel}: The cumulative distribution of expected random (unrelated) X-ray detections from forced photometry using a circular aperture of radius 30'' around 1604 optical quasar candidate positions. The black curve is the estimate obtained from Eq. 2. for a single eRASS survey. The black dashed curve shows an empirical estimate of the same number based on a random sample of positions in the surveyed field. The brown curve shows the number of expected contaminants obtained from a random sample in eRASS:4. Blue vertical lines indicate the $P_b$ values of the best detection of quasars discovered in this work in any eRASS and in eRASS:4 (if available). \textit{Lower panel}: Fraction of background sources $f_{\rm bkg}$ expected to be detected in eRASS1 from a simulation by  \citet{seppi22}.}
    \label{fig:spurious}
\end{figure}

Here $s$ and $b$ are counts within the source and background region, while r is given by the ratio of areas of these two extraction regions. 
We set the threshold of $P_b$ to be less than $0.7\%$ in the wavelength range 0.2-2.3 keV, where eROSITA is most sensitive. According to the eRASS1 simulations of \citet{seppi22}, this corresponds to a maximal fraction of $\sim 25 \%$ of spurious X-ray detections of background fluctuations. We note that this threshold is looser than the one applied by \citet[][$P_b<0.0003$]{arcodia24}, which is expected to correspond to a spurious fraction of $1\%$. However, assuming typical X-ray to optical flux ratios for Type 1 AGN \citep[e.g.][]{steffen06} $z>5.6$ quasars are expected to lie right at the sensitivity limit of the survey \citep{wolf21} and we therefore apply a more generous threshold. The fraction of spurious background fluctuation detections in eRASS, $f_\mathrm{bkg}$, predicted by \citep{seppi22} is shown as a function of $P_b$ in Fig. \ref{fig:spurious}. 

Due to the large extraction radius of the rigid source region aperture, a caveat of our detection method is the non-negligible possibility of chance alignment, i.e. the astrometric overlap of the extraction region centred on the optical coordinates of a quasar candidate and an unrelated, neighbouring eRASS source. The issue of such spurious matches in cross-survey catalogue matching has led to the development of probabilistic approaches \citep[e.g.][]{sutherland92,brusa07,budavari08,salvato18,salvato21}.  We first estimate the probability of a fully spurious association between a quasar candidate and a single-eRASS (or an eRASS:4) X-ray source based on the eROSITA catalogues. The eRASS1 source catalogue contains 1,277,477 sources. Here, we have also included sources from the supplementary catalogue, with a detection likelihood threshold $L= - \mathrm{ln}(P) \geq 5 $ in the 0.2-2.3 keV band \citep{merloni24}. Here $P$ is the probability of a source arising from a random background fluctuation \citep{brunner21}, i.e. it is directly comparable to the binomial no-source probability $P_b$ of Eq. \ref{eq:nosource}. 
We count 302,161 eRASS1 sources in the common DES, VHS, CatWISE2020 and eRASS1 footprint.
Following \citep{pineau17}, assuming that the quasar candidates and the eRASS1 sources are completely unrelated samples and using circular positional uncertainties, the total number of fully random associations is given by:

\begin{equation}
    n_\mathrm{spurious}= n_{\rm eRASS}(P_b)n_{\rm opt}\frac{\Omega_{\rm eRASS}(P_b)\Omega{\rm opt}}{\Omega},
\end{equation}

where $\Omega_{\mathrm eRASS}(P_b)=\pi \; \sigma_\mathrm{eRASS}(P_b)^2$ and $\Omega_{\mathrm opt}=\pi \; \sigma_\mathrm{opt}^2$. $ \sigma_\mathrm{eRASS}$ and $ \sigma_\mathrm{opt}$ are, respectively, the fixed source extraction radius in the eRASS map and the quasar candidates. $\Omega$ is the area of the overlap region of the eRASS and DES DR2 footprints. $n_{\rm eRASS}$ and $n_{\rm opt}$ are the number of eRASS1 sources and candidates in this overlap region.
The number of eRASS1 sources considered in the surveyed footprint, as well as typical positional uncertainties, are functions of the detection likelihood (\texttt{DET\_LIKE} in the eRASS catalogues), or equivalently the binomial no-source probability  $P_b$. We therefore bin the full eRASS1 catalogue in  \texttt{DET\_LIKE} and compute median uncertainties in each bin from the \texttt{RADEC\_ERR=$\sqrt{\sigma_{\texttt{RA}}^2 + \sigma_{\texttt{DEC}}^2 }$} values. As \texttt{RADEC\_ERR} is the quadratic sum of the 1-dimensional errors in right-ascension and declination, and we assume circular errors for our calculation ($\sigma_{\texttt{RA}} = \sigma_{\texttt{DEC}}$), we consider error radii $\sigma_{\rm eRASS}$=\texttt{RADEC\_ERR}$/\sqrt{2}$ for the eRASS1 sources \citep{pineau17}. $n_\mathrm{spurious}$ can then be calculated for each \texttt{DET\_LIKE} bin. The resulting curve of expected spurious associations is shown in Fig. \ref{fig:spurious}. At $P_b \sim 10^{-3}$, less than ten candidates out of 1604 are expected to be randomly overlapping within 30'' with an unrelated eRASS1 source. 

In addition, we have obtained an empirical estimate of chance alignments by generating 1604 random positions in the surveyed region of the sky for eRASS1 and eRASS:4. We have extracted source and background counts following the procedure detailed earlier in this section and computed $P_b$ for each of these random positions. We obtain a distribution displayed in Fig. \ref{fig:spurious}.

\begin{table}[]
\caption{Number of detected quasar candidates in each eRASS}
\centering
\begin{tabular}{@{}cccccc@{}}
\toprule
eRASS         &  1 & 2 & 3          & 4         & :4 \\ \midrule
$P_{B} < 0.007$ & 20     & 22     & 24              & 23             & 42      \\ \midrule
               &        &        & \multicolumn{2}{c}{Total unique} & 75      \\ \bottomrule
\end{tabular}
\label{tab:counts}
\end{table}

The final eRASS $z>5.6$ quasar candidate sample contains a total of 75 sources in  $3507 \, \mathrm{deg^2}$ of the sky. Table \ref{tab:counts} summarizes the number of detections in each eRASS. To each quasar candidate, we associate a probability of being randomly associated within 30'' to a background fluctuation as $P_{\rm spurious}= n_{\rm spurious}(P_b)/n_T$, where $n_T$ is the total number of quasar candidates. This value is used to prioritize targets for spectroscopic follow-up (lower values are favoured).

\section{Spectroscopic candidate confirmation}
\label{sec:three}

\begin{figure*}
    \centering
    \includegraphics[scale=0.25]{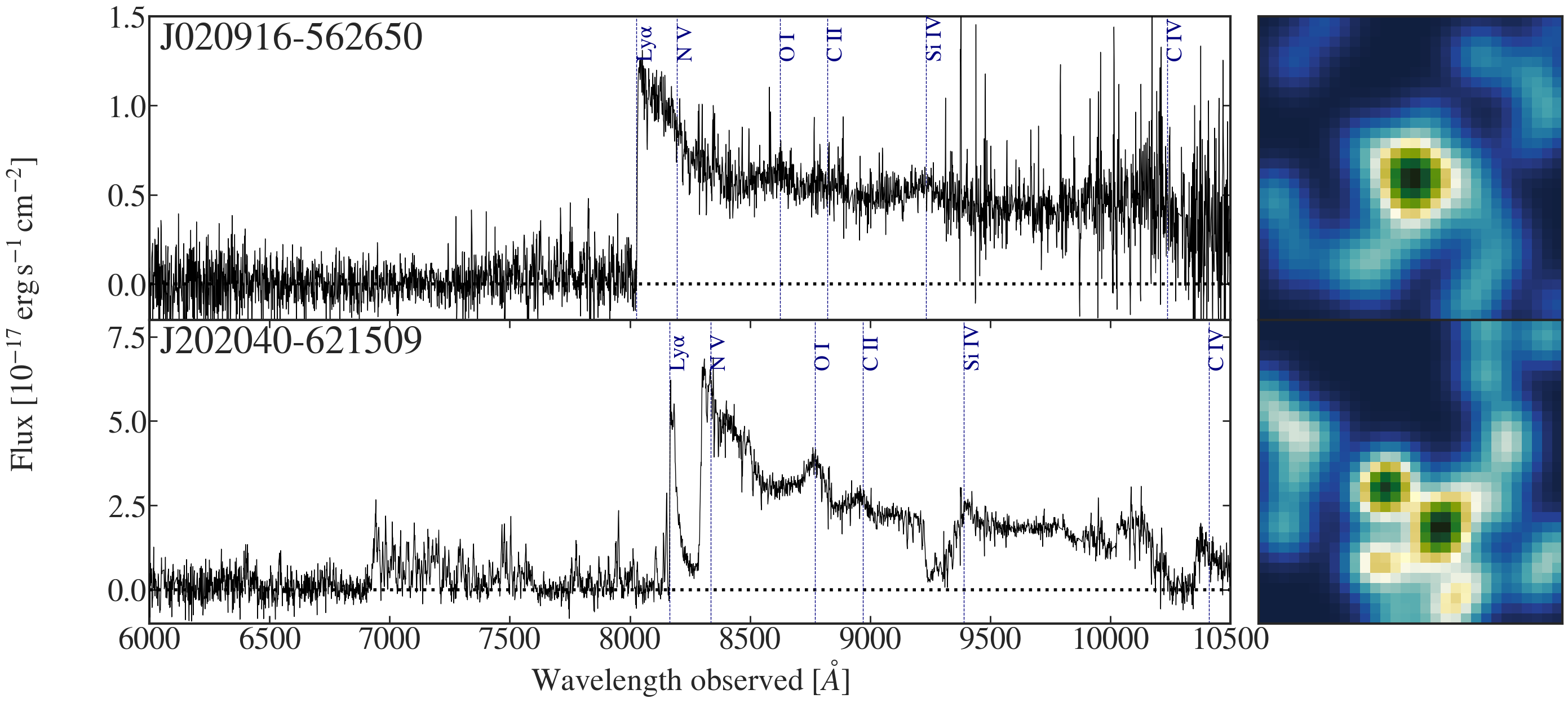}

    \caption{Spectra of newly discovered eRASS quasars (left panel) and eRASS:4 (0.2- 8 keV) postage stamps (right panel). The cutouts are $2' \times 2'$ in size and are centred on the optical position of the quasars.
    The spectrum of J202040-621509 shows prominent broad absorption features. The continuum appears strongly absorbed over the entire spectral range bluewards of Lyman $\alpha$. It appears in the eROSITA images as two distinct sources. This morphology is further investigated in Section \ref{sec:fourtwo} with a  \textit{Chandra} follow-up pointing. There we argue that the eRASS source is related to the quasar (Fig \ref{fig:bbchandra}).
    The morphology J020916-562650 is more compact.}
    \label{fig:spec}
\end{figure*}

\begin{figure}
    \centering
    \includegraphics[scale=0.37]{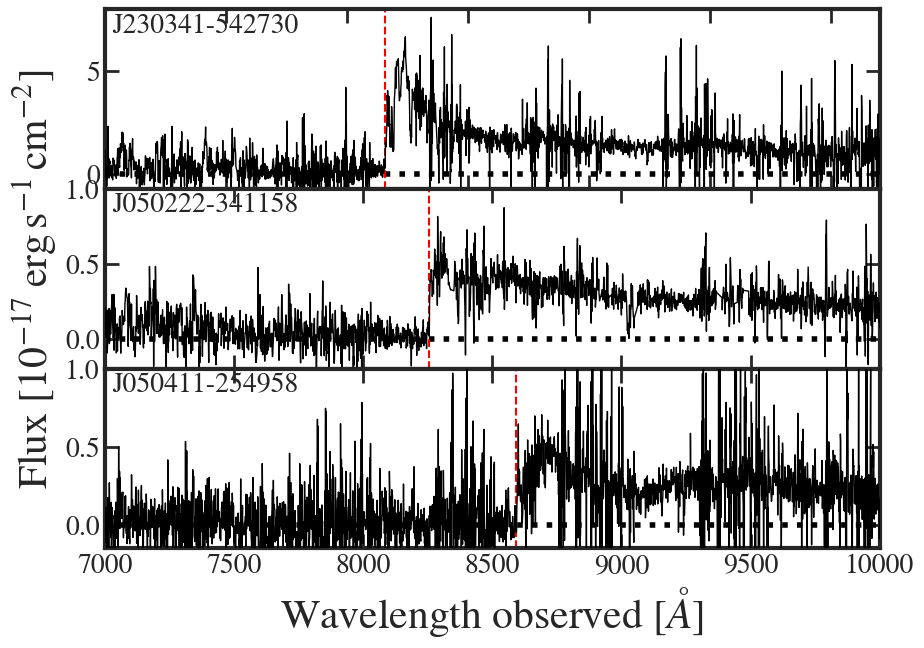}

    \caption{Spectra of three newly discovered quasars selected during the calibration of the pipeline. These objects are not robustly detected in eROSITA. The red dashed line indicates the location of the Lyman $\alpha$ break.}
    \label{fig:spec_noerass}
\end{figure}

From our 75 candidates, we obtained optical spectroscopy for two objects so far.
They were selected based on observability from the Las Campanas Observatory. In addition, as this pipeline was developed and calibrated, seven more targets were submitted for spectroscopic follow-up. All of them were selected as significant X-ray sources in eRASS data processed with an earlier version of the eROSITA reduction pipeline. In the latest processing version (020), they are not detected anymore according to the criterion defined in Section \ref{sec:erosita}. This is likely due to random fluctuations and to the fact that these objects lie at the sensitivity limit of the survey.

The targets were observed with the LDSS3-C spectrograph mounted on the \textit{Magellan}-Clay telescope at the Las Campanas Observatory (Chile). The observations took place on July 22 2021, as well as on 2021 October 12 and 26. Long-slit spectra with slit-width $1''$ were obtained using the VPH-Red grism, which spans the wavelength range 6000-10500 \AA\ at a dispersion of about 1.16 \AA/px. Total exposures varied between 1800 s to 3600 s. A summary of these observations is given in Table \ref{tab:spec}. Overscan subtractions, flat-field corrections, wavelength, and flux calibrations were performed with IRAF.

Both of the eRASS-detected quasars candidates are confirmed as high-redshift quasars\footnote{We note that shortly after the submission of the present paper, the independent discovery of quasar J020916-562650 was reported by \citet{ighina24}. The quasar J202040-621509 was also presented as quasar candidate by \citet{ighina23}. However, our spectroscopic confirmation of this quasar also pre-dates this publication.}. Their LDSS3-C discovery spectra and eRASS:4 cutouts are presented in Fig. \ref{fig:spec}. In addition, 3 of the remaining 7 targets observed with LDSS3-C in the development phase of the pipeline are also confirmed as high-redshift quasars. The spectra of these objects are presented separately in Fig. \ref{fig:spec_noerass}. The quasars are identified by the clear, abrupt, and near-total absorption of their blue continua, attributable to the Lyman $\alpha$ absorption by the increasing density of reionization-era neutral hydrogen clouds. As the unabsorbed continuum and most emission lines of the quasars are shifted out of the sampled spectral range, redshifts can only be measured from the Lyman absorption break. The measurement precision is affected by the low signal-to-noise of some of the spectra. Absorption effects redwards of Lyman $\alpha$ such as IGM damping wings \citep{rybicki79,escude98} can also severely affect Lyman $\alpha$ redshift measurements. We fit a step function with the python package \texttt{lmfit} (\texttt{StepModel() + ConstantModel()}) to the quasar spectra. We estimate the redshift and its error from the break centre by bootstrapping from the flux errors. We obtain values between $z \sim 5.6-6.1$. The remaining 4 targets all have spectra corresponding to brown dwarf templates (the coordinates of these sources are presented in Appendix \ref{append:contaminants}). We further extrapolate the restframe absolute magnitude at 1450 $\AA$, M1450, from the reddened y-band magnitudes in the DES DR2 catalogues using the spectral index $a_\nu = -0.44$ \citep{vandenberk01}. We note that because of the simultaneous DES, VHS, and CatWISE2020 selection and the respective depth of these surveys, the unabsorbed continuum slopes of the selected quasars are biased towards the red values. The choice of the extrapolation from the y-band mitigates the uncertainty due to the expected scatter in the spectral index, as its central wavelength is relatively close to 1450 $\AA$ $\times(1+z)$. M1450 measurements are listed in Table \ref{tab:spec}.

\begin{table*}[h]
 \caption{Summary of the 5 newly discovered quasars.}
\begin{tabular}{@{}cllllccc@{}}

\toprule
Source & R.A.$^{[1]}$ & Decl.$^{[1]}$ & \multicolumn{1}{c}{Observed$^{[2]}$} & \multicolumn{1}{c}{Exp. {[}s{]}$^{[3]}$} & $z_{\rm spec}$$^{[4]}$ & \texttt{mag\_auto\_z$^{[5]}$} & M1450$^{[6]}$ \\ \midrule
J202040-621509 & 305.170194 & -62.252553 & 2021-07-22 & 2400 & $5.718^{+0.001}_{-0.001}$ & $19.2345 \pm 0.0073$ & -27.32 \\
J020916-562650 & 32.320529 & -56.447344 & 2021-10-26 & 2700 & $5.606^{+0.001}_{-0.001}$ & $20.8983 \pm 0.0248$ & -25.76 \\
J050222-341158 & 75.592959 & -34.199513 & 2021-10-26 & 2700 & $5.793^{+0.001}_{-0.001}$ & $21.5331\pm 0.047$ & -25.26 \\
J230341-542730 & 345.920933 & -54.458538 & 2021-10-12 & 2400 & $5.711^{+0.004}_{-0.002}$ & $19.6143 \pm 0.0104$ & -27.06 \\
J050411-254958 & 76.049582 & -25.832959 & 2021-10-12 & 3600 & $6.072^{+0.002}_{-0.002}$ & $21.3531 \pm 0.0406$ & -25.89 \\ \bottomrule
\end{tabular}
\tablefoot{[1] Coordinates of the centroid of the optical source. [2] Observation date [3] Exposure time of the spectroscopic observation with LDSS3-C [4] Spectroscopic redshift [5] DES z-band magnitude [6] restframe 1450 $\AA$ absolute magnitude.}
\label{tab:spec}

\end{table*}

\section{Two X-ray and Radio Luminous Quasars}
\label{sec:five}

In this section, we focus on the properties of the new eROSITA quasars J202040-621509 and J020916-562650. Both are radio-bright but are intrinsically very different sources. We identify them respectively as a high-redshift jetted broad absorption line quasar (BAL QSO) and a high-redshift blazar.

\subsection{eROSITA photometry}
\label{sec:fourone}
We derived the X-ray properties of the two newly discovered and significantly eROSITA-detected quasars from the source and background spectra measured in the eRASS maps at their optical coordinates. The source and background regions are defined in Section \ref{sec:erosita} and the extraction bands are chosen to be broad (0.5 - 7 keV). Fluxes were derived in single eRASS scans, as well as in the cumulative eRASS:4 images. Extracted counts are converted to fluxes using a redshifted power-law model with fixed photon index (XSPEC model \texttt{zpowerlw}\footnote{XSPEC 12.13.1}, $\Gamma = 1.9$) and Galactic absorbing column density $N_\mathrm{H}$ from \citet{hi4pi16} with the interstellar absorption model \texttt{tbabs} \citep{wilms00}.  We opted for a power-law slope $\Gamma = 1.9$ typically observed in lower-redshift AGN populations \citep[e.g][]{liu21}. We note, however, that steeper value of the photon index, potentially related to high accretion rates, have been reported in luminous X-ray spectra of high-redshift quasars \citep[$\Gamma \sim 2.2-2.4$][]{vito19,wang20,li21,zappacosta23}. The model is scaled using the Bayesian X-ray spectral analysis code \texttt{BXA} \citep{buchner14}. The spectra are rebinned with the \texttt{ftgrouppha} function of the \texttt{FTOOLS}  package\footnote{https://heasarc.gsfc.nasa.gov/ftools/}, ensuring at least one bin in each count. The data are analysed using the Cash statistic \citep[w-stat,][]{cash79,wachter79}. We obtain median values, as well as the 15.9th and 84.1th percentiles\footnote{These percentiles correspond to the $\mu \pm 1 \sigma$ locations of the normal distribution.} of the posterior flux chains. For the single epoch photometry, we report the flux in the eRASS for which $P_b$ reaches its minimum. We only consider epochs with $P_b<0.007$. Quasar J202040-621509 is only detected in eRASS1 ($P_b=4.4\times 10^{-4}$), while J020916-56265 is detected in eRASS2, eRASS3, and eRASS4 (with a minimum of $P_b=2.2\times 10^{-7}$). The cumulative photometry is computed from the stacked survey eRASS:4 (the deepest, homogeneous, and complete cumulative eRASS survey). We summarize the X-ray properties of the discovered quasars in Table \ref{tab:xray}. 
We derived restframe hard X-ray luminosities $L_{2-10 \mathrm{keV}}$ assuming the above AGN spectral model. Extrapolating from the dereddened y-band DES magnitudes assuming the spectral slope of $\alpha=-0.44$ \citep{vandenberk01} we computed an estimate of the restframe monochromatic luminosity at 2500 $\AA$, $L_{2500}$. We compute the two-point spectral index $\alpha_{OX}$ connecting the optical and X-ray part of the quasar SED as: 
\begin{equation}
    \alpha_{\mathrm{OX}} = 0.384 \times \mathrm{log}\, (L_{2 \, \rm keV}/L_{2500}), 
    \label{eq:aox}
\end{equation}
where $L_{2500}$ is the luminosity at $2 \, \mathrm{keV}$.

\begin{figure*}
    \centering
    \includegraphics[scale=0.192]{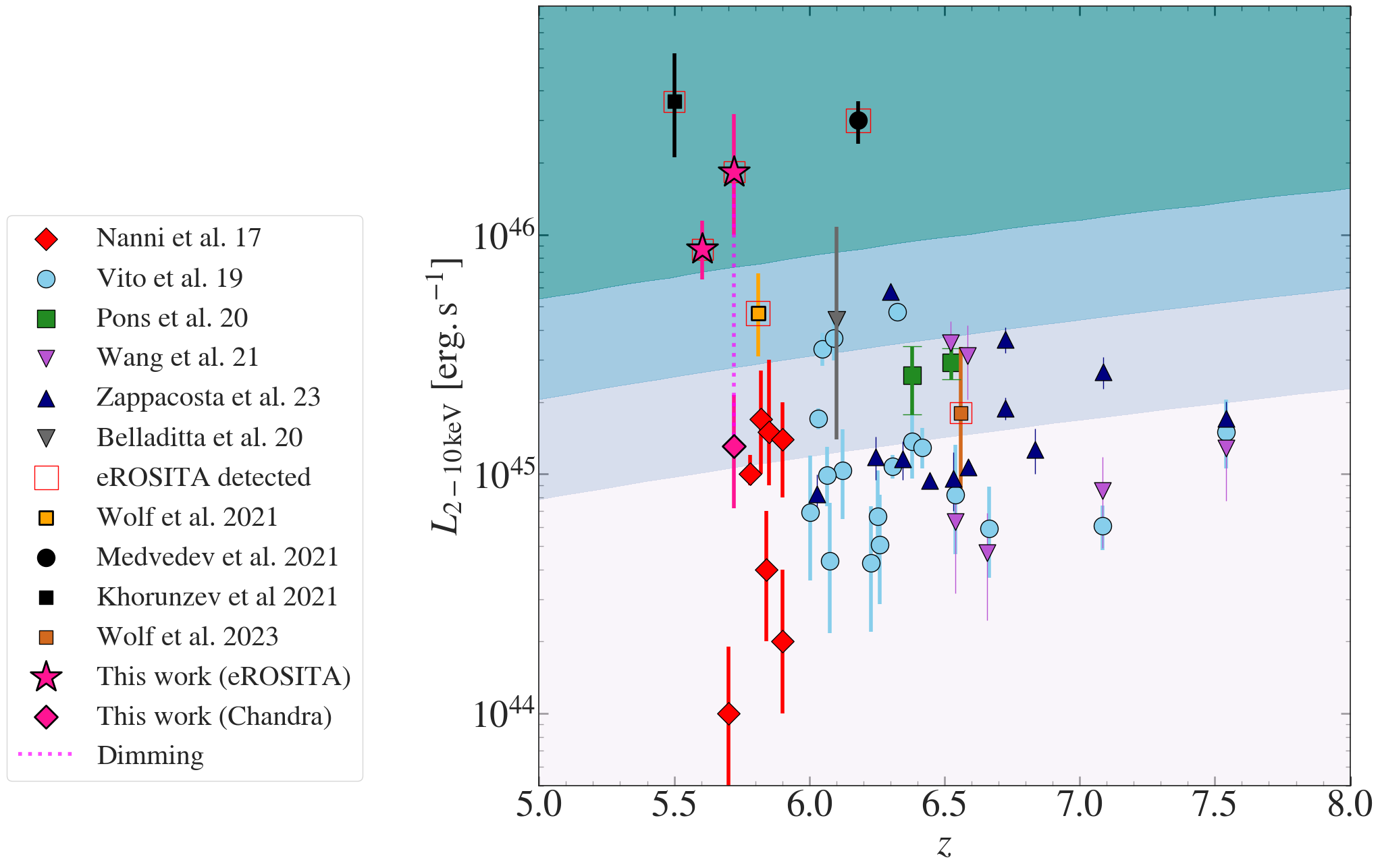}
    \includegraphics[scale=0.192]{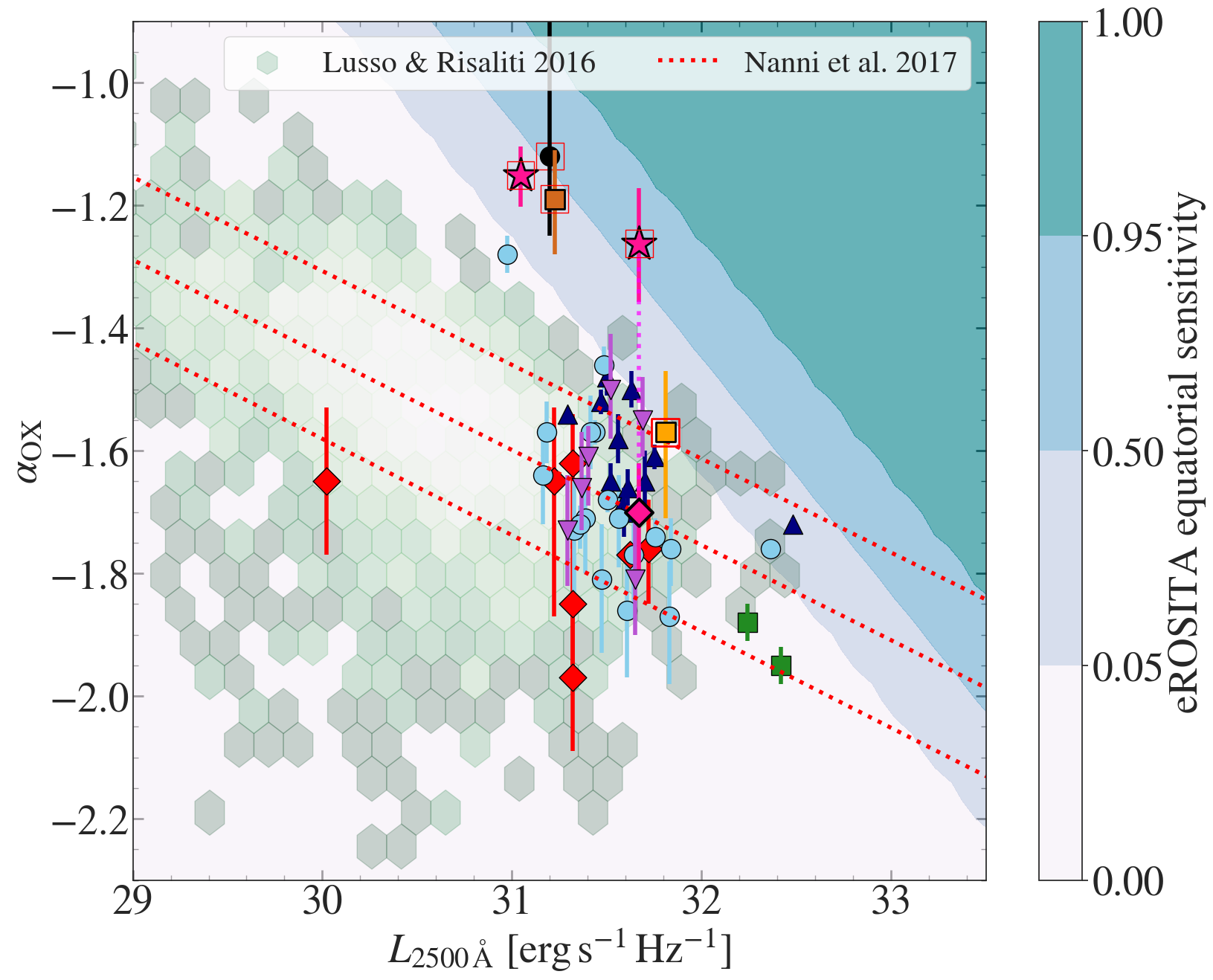}

    \caption{\textit{Left panel}: X-ray luminosity redshift distribution of X-ray luminous quasars from literature \citep{nanni17,vito19,pons20,wang20,medvedev20,belladitta20,wolf21,wolf22,zappacosta23} spanning a redshift range $5.5<z<7.5$ and a luminosity range $44<\mathrm{log} \, L_{2-10 \, \mathrm{keV}}/\mathrm{(erg/s)}<46.5$. The two newly discovered quasars (displayed as pink stars). The colour gradient shows the sensitivity to a fiducial absorbed power-law with Galactic absorption of the final cumulative eRASS in the equatorial region (in fractional of sensitive area). The quasars discovered in this work lie at the luminous end of the quasar population in the early universe. We show the luminosity derived from the \textit{Chandra} follow-up observation of J202040-621509. Between the eRASS1 and the \textit{Chandra} observation, its luminosity has decreased by an order of magnitude. \textit{Right panel}:  $\alpha_{OX}$-L2500 distribution of the same sample of sources. The 1$\sigma$ confidence interval of relation derived by \citet{nanni17} is shown by red dotted line. The hexagonal pattern shows a lower redshift AGN sample by \citet{lusso16}. The three quasars X-ray detected and newly discovered are over-luminous in the X-ray wavebands. Following its dimming observed in the recent \textit{Chandra} observation, J202040-621509 is perfectly consistent $\alpha_{OX}$-L2500 relation. } 
    \label{fig:lxaox}

\end{figure*}

In Fig. \ref{fig:lxaox}, left panel, we present the $L_{2-10 \rm keV}$-redshift plane
X-ray detected quasars in a redshift range $5.5<z<7.5$ and a luminosity range $44<\mathrm{log} \, L_{2-10 \, \mathrm{keV}}/\mathrm{(erg/s)}<46.5$ \citep{nanni17,vito19,pons20,medvedev20,khorunzev21,wang20,wolf21,wolf22,zappacosta23}. The average eROSITA sensitivity in units of fractional sensitive area is shown as a colour map (values for a typical equatorial eROSITA field). It was computed by forward-modelling the sensitivity of the instrument to a fiducial AGN, described here by an absorbed power-law with $\Gamma = 2$ and Galactic absorption, redshifted to $z= 6$ \citep[see][]{wolf21,wolf22}. The quasars J202040-621509 and J020916-562650 are among the most X-ray luminous quasars known to date. 
The spectral index $\alpha_{OX}$ is known to anti-correlate steeply with rest-frame UV luminosity L2500 for Type 1 AGN \citep[e.g.][]{steffen06,just07,lusso16}. We show the relation measured by \citet{nanni17} for a sample of $z>5.5$ quasars. In the right panel, we compare the $\alpha_{OX}$-L2500 of the newly discovered quasars with the sample X-ray detected ones from the literature. The $\alpha_{OX}$ values of J202040-621509, J020916-562650, and J230341-542730 significantly deviate from the measured relation. Their location indicates that they are X-ray over-luminous.

\begin{table*}[]
\caption{eROSITA X-ray properties from aperture photometry}
\begin{tabular}{@{}cccccccc@{}}
\toprule
Source & eRASS$^{[1]}$ & eRASS:4?$^{[2]}$ & \begin{tabular}[c]{@{}c@{}}Net cts.$^{[3]}$\\ \end{tabular} & \begin{tabular}[c]{@{}c@{}}$F_{\rm 0.5-7 \, kev, s}$ $^{[4]}$\\ $\mathrm{[erg\, cm^{-2} s^{-1}]}$\end{tabular} & \begin{tabular}[c]{@{}c@{}}$F_{\rm 0.5-7 kev, c}$ $^{[5]}$\\ $\mathrm{[erg\, cm^{-2} s^{-1}]}$\end{tabular} & \begin{tabular}[c]{@{}c@{}}$L_{\rm 2-10 \, keV}$ $^{[6]}$ \\ $\mathrm{[erg\, s^{-1}]}$\end{tabular} & $\alpha_{OX}^{[7]}$ \\ \midrule
J202040-621509 & 1 & No & $9.5 \pm 3.9$ & $7.82^{+11.06}_{-4.47} \times 10^{-14}$ & unconst. & $1.83^{+1.35}_{-0.83}\times 10^{46}$* & $-1.27^{+0.09}_{-0.10}$ \\
J020916-562650 & 2,3,4 & Yes & $21.5 \pm 5.3$ & $4.99^{+3.71}_{-2.02}\times 10^{-14}$ & $2.57^{+1.35}_{-0.80}\times 10^{-14}$ & $8.70^{+2.71}_{-2.18}\times 10^{45}$ & $-1.15^{+0.05}_{-0.05}$ \\ \bottomrule
\end{tabular}
 \tablefoot{[1] eRASS in which the quasar is detected. The bold font indicates the eRASS in which $P_b$ is minimal. [2] Was the source detected in eRASS:4 ? [3] Net counts in the 0.2-2.3 keV eRASS:4 images [4] maximal single eRASS flux in the 0.5-keV band. [5] eRASS:4 flux [6] Hard restframe X-ray luminosity. $*$We note that the luminosity of J202040-621509 is only based on its eRASS1 detection, while for J020916-562650 we used the cumulative data from eRASS:4. [7] Spectral index connecting X-ray to optical SED. }
\label{tab:xray}

\end{table*}

\subsection{eRASS J020916-562650: A blazar at $z=5.6$ }

\subsubsection{Radio properties}

J020916$-$562650 is clearly detected at 0.887 and 1.37 GHz in the Rapid ASKAP Continuum Survey (RACS, \citealt{mcconnell}; \citealt{hale21}), at 0.887 in the ASKAP Variables and Slow Transients
survey (VAST, \citealt{murphy13}) and at 0.943 GHz in the Evolutionary Map of the Universe survey (EMU, Norris et al. 2021). 
To quantify the flux densities at these frequencies, we performed a single Gaussian fit using the task IMFIT of the Common Astronomy Software Applications
package (CASA, \citealt{mcmullin07}) on the images retrieved from the CASDA\footnote{The CSIRO ASKAP Science Data Archive, \url{https://data.csiro.au/domain/casda}} website. 
They are reported in Appendix \ref{append:radio}, Table. \ref{J0209radio_fluxes}. 
By assuming <a single power-law for the continuum radio emission (S$_\nu$ $\propto$ $\nu^{-\alpha}$) we estimated a radio spectral index ($\alpha_r$, see Fig. \ref{radiospecs}) of  0.14$\pm$0.10, which classifies it as a Flat Spectrum Radio Quasar (FSRQ, e.g. Urry \& Padovani 1995).
The two flux densities estimated from the VAST images show a hint of variability on a time scale of $\sim$2 months (in the observed frame, i.e., $\sim$9 days in the rest frame). 
Then we computed the values of radio-loudness (R), which quantify how powerful is the radio (i.e. synchrotron) emission with respect to the optical/UV (i.e. accretion disk) one. 
We used the definition of Kellermann (1989): R= S$_{5 GHz}$/S$_{4400\AA}$. 
We derived the flux density at 5~GHz rest frame by using the available observed flux densities (0.87, 0.94 and 1.3 GHz) and the estimated radio spectral index. 
The value of the flux density at 4400\AA\ rest frame was derived from the z-band magnitude assuming the optical spectral index of $\alpha_{\nu}$ = 0.44 (Vanden Berk et al. 2001). 
We obtained a mean R of 750$\pm$170 (Log(R) = 2.87).
The high value of radio loudness, the flat radio spectral index and tentative variability indicate that this source can be classified as a blazar, i.e. a radio-loud AGN with the relativistic jet pointed towards the Earth (e.g., Urry \& Padovani 1995).

\subsubsection{X-ray spectral analysis}

\begin{figure}
    \includegraphics[scale=0.65]{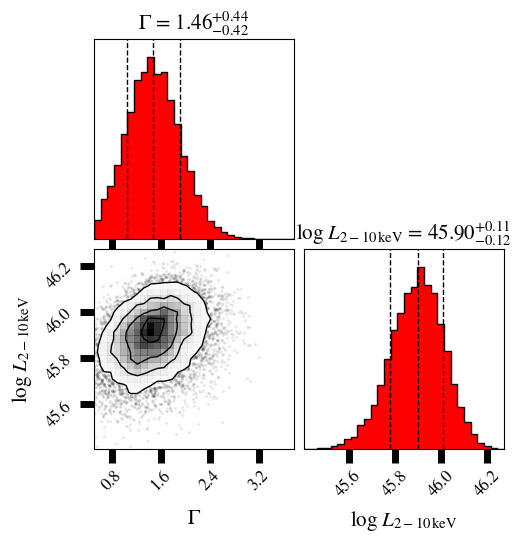}
    \caption{Posteriors on the photon index $\Gamma$ and the X-ray luminosity  $L_{\rm 2-10\, keV}$ from the X-ray spectral analysis performed on the eRASS:4 spectrum of J020916-562650}
    \label{fig:xspec}
\end{figure}

To further investigate the potential blazar nature of the source, we attempt a fit of the eRASS:4 spectrum of J020916$-$562650, by modelling it once again by a power-law with Galactic absorption (NH = $2.18 \times 10^{20} \, \mathrm{cm^{-2}}$, \citealt{hi4pi16}): \texttt{zpowerlw}$\times$\texttt{tbabs}. The photon index $\Gamma$ is left free to vary in the fit. We stress that the low-counting statistics ($\sim 25$ net counts in the source region), significantly limit our ability to obtain a well-constrained X-ray spectral shape for this source. The analysis is performed with \texttt{BXA} \citep{buchner14}. We define wide and uninformative priors for both the normalization and the slope of the power-law: $\Gamma=0.5 -5$ (uniform prior) and $N = 1^{-6}-1$ (Jeffrey's prior). The eRASS:4 source and background spectra are extracted in regions as defined in Section \ref{sec:erosita}. We have used the \texttt{srctool} function of the software package \texttt{eSASS} (user version
 eSASSusers\_240410) to obtain these spectra.
We use the redshift of the source measured in Section \ref{sec:three}.
The posterior distributions of the photon index and the hard X-ray luminosity  $L_{\rm 2-10\, keV}$ we obtain are presented in Fig. \ref{fig:xspec}. Due to the poor counting statistics, the photon index is not well constrained: $\Gamma = 1.46^{+0.44}_{-0.42}$. Flatter values than typical AGN are, however, preferred \citep[e.g.][]{nandra94,reeves97,ishibashi10}. This flatter photon index value is consistent with values typically observed in the restframe hard X-ray spectra of lower redshift blazars \citep[e.g.][]{langejahn20}. Deeper X-ray data will be required to confirm the flatness of the X-ray spectrum.

The luminosity we obtain with this photon index is log $L_{\rm 2-10\, keV}/[\mathrm{erg\,s^{-1}}]=45.89^{+0.11}_{-0.12}$. From Eq. \ref{eq:aox} and this luminosity estimate, we obtain a spectral index $\alpha_{OX}=-1.15^{+ 0.01}_{-0.01}$. This relatively flat value, given the quasar's optical monochromatic luminosity, indicates that the quasar is X-ray over-luminous with respect to typical Type 1 AGN. Such a flat value is typical for blazars \citep[e.g.][]{donnato01}.

\subsubsection{A blazar SED}

\begin{figure}
    \includegraphics[scale=0.42]{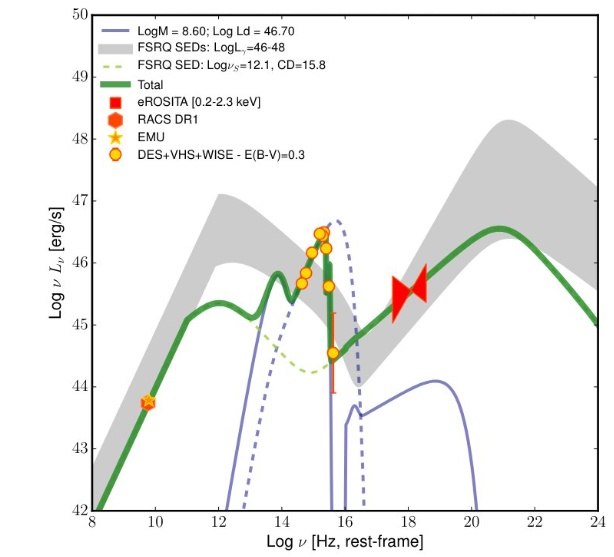}
    \caption{The broadband rest frame SED of J020916$-$562650. A range of typical FSRQ SEDs ($L_\gamma\sim10^{46-48}$erg/s, grey shaded region) and a more tailored blazar analytical model (solid green line) are overplotted to the data. The radio and X-ray emission of the source are consistent with 
    the typical blazar double hump SED. Specifically, the X-ray emission significantly overestimates the expected X-ray coronal emission, extrapolated by the source accretion emission that dominates the optical/UV restframe photometry (yellow dots). The thermal emission is modelled here with a standard accretion disk emission, IR torus, and X-ray Corona powered by a black hole with the mass of log $M_{\rm BH}/M_\odot=8.60$ (blue solid and dashed lines, the latter showing the unabsorbed accretion disk SED).}
    \label{fig:sed_blazar}
\end{figure}

In the previous section, we were able to constrain $\Gamma$ and $\alpha_{OX}$ values for J020916$-$562650 that are typical for blazars. We further have shown that the radio morphology, spectral shape, and variability of the associated radio source are consistent with the hypothesis that the source is jetted and that its jet aligns with the line of sight. 
Fig.\ \ref{fig:sed_blazar} shows the broadband rest frame SED of the source, with data from radio to the X-rays. 
The observed data are compared with the analytic approximation of a typical FSRQ SED as developed by \citet{ghisellini17}, combined with the thermal accretion emission from a standard geometrically thin, optically thick $\alpha$-disk \citep{shakura73}, with relative X-ray Corona and dusty IR torus. 
Such thermal emission clearly dominates the rest-frame NIR-optical-UV energy range in this case, thus the DES, VHS and CATWISE2020 photometric points have been fitted with a \citet{shakura73} model, following a method already introduced by \citet{sbarrato12,calderone13,ghisselini15,belladitta19} and \citet{belladitta22}. 
Following this approach, we can conclude that the central engine is likely fast accreting. Its emission is consistent with a log $M_{\rm BH}/M_\odot=8.60$ black hole, accreting at the Eddington limit. 
The radio and X-ray data points of J020916-562650 are instead clearly dominated by the emission from a relativistic jet, aligned closely to our line of sight. 
Even with a sparse sampling of the broad-band SED, the source shows the typical double hump SED profile of blazars: the hard and intense X-ray luminosity can only be justified by an inverse Compton emission, being strongly overluminous with respect to the expected Corona emission. The significant radio luminosity further confirms the blazar interpretation, tracing an intense synchrotron emission.

\subsubsection{AGN demographics from a single blazar discovery}

\citet{volonteri10b} and \citet{ghisselini10} developed the geometric argument that for each blazar, i.e. line-of-sight-aligned jet (viewing angle below the critical angle $<1/\Gamma_\mathrm{Lorentz}$, where $\Gamma_\mathrm{Lorentz}$ the bulk Lorentz factor), there must exist $2 \times \Gamma_\mathrm{Lorentz}^2$ parent jets in quasars pointed in any other direction. Following the above argument, \citet{belladitta20} derived a lower limit on the space density of AGN similar in the same redshift-magnitude bin than PSO J030947.49+271757.31, the most distant blazar known to date. Assuming a typical value for the bulk Lorentz factor ($\Gamma_\mathrm{Lorentz} = 10$, e.g. \citealt{ghisselini14}) and considering a redshift interval $z=5.6-6.5$ over the surveyed area of $3507 \, \mathrm{deg^2}$,  we obtain a number density of quasars at least as bright as J020916-562650 (M1450$<25.76$) of $6.89^{15.86}_{-5.70}\, \mathrm{Gpc^{-3}}$. The Poisson confidence limits are estimated according to \citet{gehrels86}. This lower limit on the space density of the parent jetted quasar population is consistent with \citet{belladitta19} at similar absolute magnitudes ($1.10^{+2.53}_{-0.91} \, \mathrm{Gpc^{-3}}$ at M1450$\leq 25.1$ and within $z=5.5-6.5$). \citet{belladitta19} compare their result to the integrated quasar luminosity function by \citet{matsuoka18lf} ($3.64^{+0.09}_{-0.03} \, \mathrm{Gpc^{-3}}$ at M1450$<-25.1$), concluding that $30\%$ of AGN at $z=5.5-6.5$ are radio loud.

Integrating the quasar luminosity function of \citep{schindler23} over the same redshift range down to M1450$<25.7$ yields a space density $\sim 1.32 \, \mathrm{Gpc^{-3}}$. The density of luminous jetted quasars derived from the discovery of J020916-562650 is consistent with the previous density measurements at these redshifts and luminosities by \citet{jiang16,matsuoka18lf,schindler23}. We note, however, that the argumentation above requires that all of the quasars we account for are similarly radio-loud and jetted. This suggests that either most of the jets of the quasars discovered so far in this redshift range have not yet been detected in radio surveys or that a largely hidden population of potentially obscured quasars have eluded dedicated surveys, as the radio-loud fraction of quasars is not expected to exceed $\sim 30 \%$ at these redshifts \citep{banados15, belladitta19}. A similar conclusion was reached by \citet{caccianiga24}, who compared the number of radio detected quasars to the number of blazars. A possible source of error in our estimate is the value we have assumed for the bulk Lorentz factor or the $\Gamma_\mathrm{Lorentz}^2$ dependence on the number density of parent jets. \citet{lister19} argue that the simple geometric argument we have made use of fails to account for biases arising from flux-limited sampling. These authors show through Monte Carlo simulations that the evolution of the fraction of jetted quasars to blazars (brighter than 1.5 Jy at 15 GHz) as a function of $\Gamma$ is shallower than the assumed relation $\propto \Gamma_\mathrm{Lorentz}^2$ evolution at $\Gamma_\mathrm{Lorentz}>15$. We note, however, that \citet{lister19} a derive steep relative evolution of blazar jets and their parent population at $\Gamma_\mathrm{Lorentz}<15$, resulting in $N_\mathrm{parent} \sim 10^3 \times N_\mathrm{>1.5 ¸\, Jy}$.
Due to the large surveyed area and the rarity of AGN, the Poisson counting uncertainties fully dominate cosmic variance \citep[e.g.,][]{trenti08, dai15}. Our results underline the necessity of systematic, multi-wavelength searches for high-redshift blazars to better constrain AGN demographics in the poorly sampled luminous regime.

\subsection{eRASS J202040-621509: X-ray dimming in a radio-bright $z=5.7$ BAL quasar ?}

\subsubsection{Confirming the X-ray detection with \textit{Chandra} ACIS-S}
\label{sec:fourtwo}

\begin{figure*}
    \centering
    \includegraphics[scale=0.4]{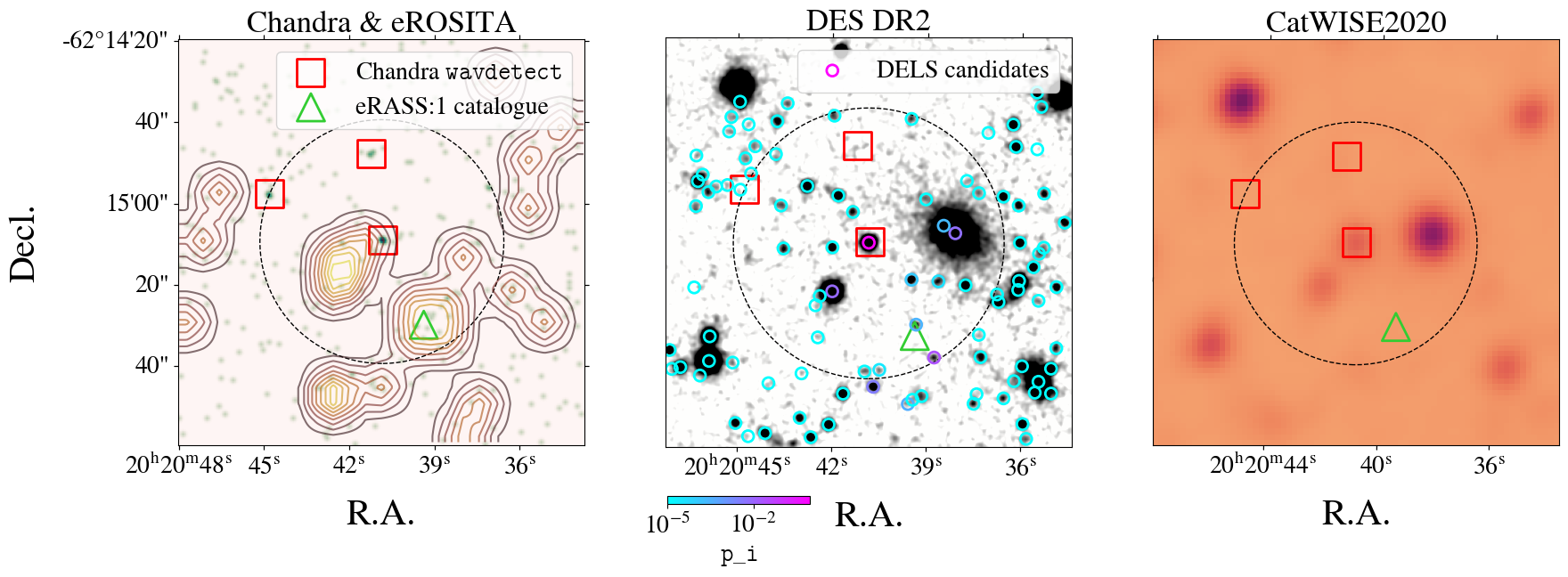}
    \caption{All images are centred on the optical position of the new quasar J202040-621509 and have sizes 100'' x100''. \textit{Left panel}: Broadband (0.5-7 keV) \textit{Chandra} image. \textit{Chandra} sources detected with 30'' (black, dashed circle) of the quasar coordinates with \texttt{wavdetect} are marked in red. The centroid of 1eRASS J202039.8-621525 catalogue detection is marked by a green triangle. The contours were obtained from the smoothed eRASS:4 0.5 - 7 keV image. \textit{Central panel:} DES DR2 z-band image. Circular coloured markers denote counterpart candidates for 1eRASS J202039.8-621525 from the DESI Legacy Survey catalogue. They are colour-coded according to their relative probability \texttt{p\_i} of being the best counterpart. The DESI Legacy Survey source with the highest \texttt{p\_i} is spatially co-incident with the quasar coordinates. \textit{Right panel:}  CatWISE 2020 W1 image. Among the \textit{Chandra} and eROSITA X-ray detections, only the quasar is unambiguously detected. }
    \label{fig:bbchandra}
\end{figure*}

To increase counting statistics and constrain the X-ray spectral shape of J202040-621509, we obtained an additional \textit{Chandra} ACIS-S pointing.
 J202040-621509 was observed with \textit{Chandra} ACIS-S on September 23, 2023 as Cycle 23 GTO target (PI: Predehl, Observer: Wolf, obsid: 26040). The total exposure time was 29.58 ks. We analyzed the data with the X-ray ray data analysis software \texttt{ciao 4.15}. We reduced the data with the latest pipeline and calibration database version. To detect sources in the image, we ran the Mexican-hat wavelet algorithm \texttt{wavdetect} on the obtained broad band images (0.5-7 keV). We use the default detection parameters (\texttt{sighthresh}$=10^{-6}$ and \texttt{bkgsigthresh}$=0.001$). A source is significantly detected within $0.3''$ of the quasar with $9.9 \pm 3.2$ net counts. Given the astrometry, the X-ray source is unequivocally associated with the quasar. We refer to this source as \textit{ch1} henceforth. Two more sources, which we label \textit{ch2} and \textit{ch3}, are detected at a lower significance level within $30''$ of the quasar. They are located to the north and northeast of the quasar and do not overlap with the eROSITA X-ray centroids to the south. The eRASS:4 and \textit{Chandra} extraction images are overlayed in Fig. \ref{fig:bbchandra} (left panel). From the eRASS:4 image, we count 7 photons in the source region, making the source position and morphology highly uncertain. An eRASS1 catalogue source, 1eRASS J202039.8-621525, is located within 17.71'' of the optical coordinates of the quasar (green triangle). To help us determine the source of the eROSITA signal, we performed a cross-match between the eRASS1 catalogue and the Dark Energy Spectroscopic Instrument (DESI) Legacy Imaging Survey (\citealt{dey19}) with the probabilistic cross-matching algorithm \texttt{NWAY} \citep{salvato18}. Our method combines optical/IR photometric priors derived from a truth sample of X-ray sources from the Fourth XMM-Newton Serendipitous catalogue \citep[4XMM,][]{webb20} with the astrometric configuration (distances, counterpart candidate surface density and positional errors). The procedure is presented in detail by \citet{salvato21}. In our setup, all DESI Legacy Survey sources within a radius of 1' around an eRASS1 source are considered as potential candidates. 151 such DESI Legacy Survey candidates are found. We show a DES DR2 z-band image showing the respective coordinates of the centroid of eRASS1 J202040-621509, \textit{ch1}, \textit{ch2}, \textit{ch3} and the DESI Legacy Survey neighbouring candidates in Fig. \ref{fig:bbchandra} (central panel). \texttt{NWAY} outputs the matching statistics \texttt{p\_any} and \texttt{p\_i}, which are respectively the probability that any of the candidates is matched to the X-ray source and the relative probability among candidates. In the case of 1eRASS J202039.8-621525 (\texttt{DETUID:em01\_306153\_020\_ML00171\_002\_c010}, DET\_LIKE\_0 = 9.95), we obtain \texttt{p\_any}$=0.012$. The best matching source is a DESI Legacy Survey source matched within $0.16''$ with the quasar DES coordinates ($\texttt{p\_i}=0.97$). We note that closer DESI Legacy Survey counterpart candidates are down-weighted by the custom photometric priors we employ, as their colours do not correspond to that of typical X-ray sources. Preliminary calibrations using a random test sample show that a \texttt{p\_any}$=0.012$ roughly corresponds to a purity of $\sim 45 \%$ at a completeness of $\sim 100 \%$ in the retrieved counterpart samples. From this exercise, we cannot conclude with certainty that 1eRASS J202039.8-621525 is associated with the quasar, but we can exclude the other DESI Legacy Survey sources in the field as the source of the X-ray emission to a high degree of certainty ($\texttt{p\_i}<0.02$). Many of the DESI Legacy Survey sources appear to be duplicate our spurious. The offset \textit{Chandra} source \textit{ch2} appears matched to a catalogue DESI Legacy Survey source, while \textit{ch3} has no obvious counterpart. Performing forced photometry on background-subtracted DES images within circular source apertures of radii equal to the semi-major axis of the \textit{Chandra} error ellipses (respectively 2.5'' and 1.5''). We obtain a S/N$<3$ in all DES bands for both sources, suggesting that these two \textit{Chandra} sources are not matched in DES.

 We extracted fluxes for the three \textit{Chandra} sources using the \texttt{srcflux} routine of the \texttt{ciao} software package. For all three sources, we assumed a fiducial power-law model  \texttt{zpowerlw} with Galactic absorption and fixed photon index ($\Gamma=1.9$). We report the broad-band photometry for these three detections in Table \ref{tab:acis}. \textit{ch1} is the most significant detection among the three sources near the quasar. We note that no counts are registered in the ultrasoft band (0.2 - 0.5 keV) and the soft band (0.5-1 keV) for \textit{ch1}. For this source, we measure $5.85 \pm 2.45$ net counts in the medium band (1 - 2 keV) and $4 \pm 2$ counts in the hard band (2-7 keV). This hard photometry can be indicative of a significant level of obscuration, as no counts can be measured in the restframe 1.34 - 6.7 keV energy of the quasar SED.

 In summary, we have presented evidence that supports that 1eRASS J202039.8-621525 is associated with the quasar emission, as (A) the \textit{Chandra} sources \textit{ch2} and \textit{ch3} are marginally detected and do not overlap with the eRASS contours. (B) These sources are also not significantly detected in DES and CatWISE2020. (C) Among 151 potential DESI Legacy Survey candidates within 1' of 1eRASS J202039.8-621525, the candidate with the highest relative probability \texttt{p\_i} of being associated to the X-ray source is matched at the sub-arcsecond level with quasar coordinates. (D) No other \textit{Chandra} source is detected to the southwest of the quasar, close to eROSITA events. 
While we cannot exclude the possibility of a transient phenomenon in a foreground galaxy, we conclude that it is likely that the bright eRASS1 source is associated with the quasar.

\begin{table}[]
\caption{\textit{Chandra} ACIS-S photometry for three sources detected within 30'' of the quasar.}
\begin{tabular}{@{}lllll@{}}
\toprule
\multicolumn{1}{c}{ID} & \multicolumn{1}{c}{\begin{tabular}[c]{@{}c@{}}Sep.\\ {[}arcsec{]}\end{tabular}} & \multicolumn{1}{c}{\begin{tabular}[c]{@{}c@{}}Net counts\\ 0.5-7.0 keV\end{tabular}} & \multicolumn{1}{c}{\begin{tabular}[c]{@{}c@{}}Significance\\ $[\sigma]$\end{tabular}} & \multicolumn{1}{c}{\begin{tabular}[c]{@{}c@{}}$F_\mathrm{0.5-7.0 keV}$\\ {[}$\mathrm{erg\, s^{-1}\, cm^{-2}}${]}\end{tabular}} \\ \midrule
\multicolumn{1}{c}{\textit{ch1}} & \multicolumn{1}{c}{0.30} & \multicolumn{1}{c}{$9.9 \pm 3.2$} & \multicolumn{1}{c}{3.06} & \multicolumn{1}{c}{$5.83^{+3.70}_{-2.63}$} \\
\multicolumn{1}{c}{\textit{ch2}} & \multicolumn{1}{c}{21.57} & \multicolumn{1}{c}{$4.7 \pm 2.2$} & \multicolumn{1}{c}{2.00} & \multicolumn{1}{c}{$2.86^{+2.98}_{-1.82}$} \\
\multicolumn{1}{c}{\textit{ch3}} & \multicolumn{1}{c}{29.96} & \multicolumn{1}{c}{$6.6 \pm 2.7$} & \multicolumn{1}{c}{2.40} & \multicolumn{1}{c}{$3.80^{+3.20}_{-2.09}$} \\
\bottomrule

\end{tabular}
\tablefoot{[1] \textit{Chandra} source [2] The separation in arcsec between the optical position and the \textit{Chandra} source. [3] The net counts in the band 0.5-7 keV  [4] The srcflux-based significance of the detection based on source and background counts. [5] The flux in observed 0.5-7 keV.}

\label{tab:acis}
\end{table}

\subsubsection{Long-term X-ray variability}
\label{sec:xvar}
As presented in Table \ref{tab:xray}
, J202040-621509 is only detected in eRASS1, but not in any subsequent scan. 

The eROSITA image presented in Fig. \ref{fig:bbchandra} displays two distinct lobes, which extend to the south and southwest of the coordinates of the quasar. One of these lobes coincides astrometrically with the quasar. In the \textit{Chandra} image, three sources were detected with the \texttt{wavdetect} algorithm. One of these sources is matched at the sub-arcsecond level to the quasar's coordinates. The two other sources are detected at a lower significance level to the north and the northwest of the quasar. Their location does not match the eRASS contours, and they are associated with DES sources within 1''. Assuming that the eRASS1 photometry is solely associated with the quasar, the eRASS and \textit{Chandra} photometry presented respectively in Section \ref{sec:fourone} and \ref{sec:fourtwo}, imply a decrease in the measured 0.5 - 7 keV flux of the source of about an order of magnitude from $F_{0.5-7 \rm keV}=7.82^{+11.06}_{-4.47} \times 10^{-14}\mathrm{erg\,s^{-1}cm^{-2}}$ to $F_{0.5-7 \rm keV}=5.83^{+3.70}_{-2.63} \times 10^{-15} \mathrm{erg\,s^{-1}cm^{-2}}$. We note that the flux decrease is only significant at the $\sim 2\sigma$ level due to the low number of net source counts in eRASS1. The two observations are separated by 3.5 years in the observed frame, corresponding to 6 months in the quasar restframe. 

\begin{figure}
    \includegraphics[scale=0.38]{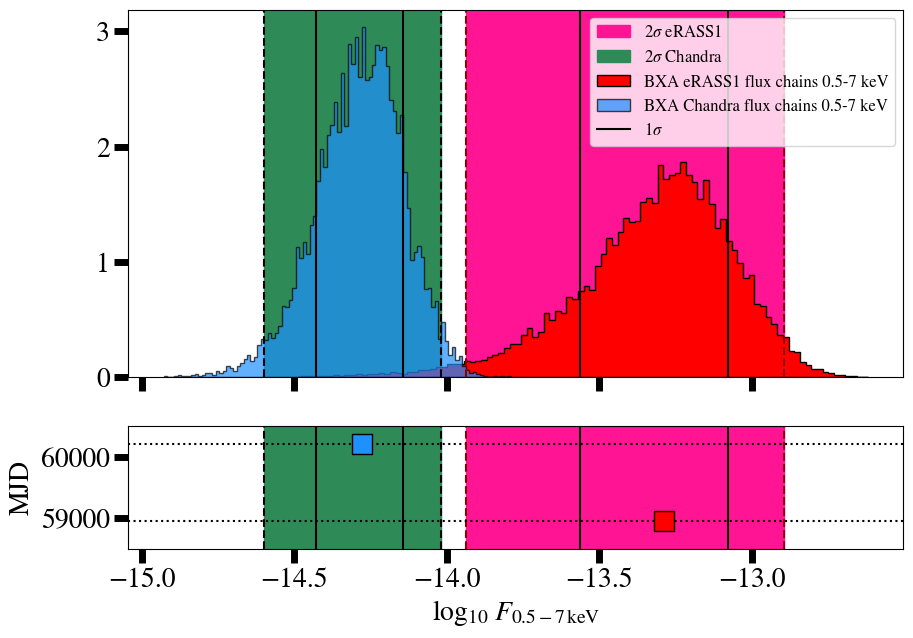}
    \caption{Distribution of \texttt{BXA} $F_{0.5-7 \rm keV}$ fluxes for the eRASS 1 (in blue) and \textit{Chandra} observations (red) of J202040-621509. The vertical black lines show the locations of the 15.9-th and 84.1-th percentiles of the distributions. The green and pink regions span the 2.3-th and 97.7-th percentiles of the distributions. 
    The lower panel shows the time separation in days between the two observations. We measure a decrease in the source flux between the two epochs significant at the $\sim 2 \sigma$ level.}
    \label{fig:dimming}
\end{figure}

Fig. \ref{fig:dimming} shows the flux-decrease in the broad 0.5-7 keV band. The BXA flux posteriors for the eRASS1 and Chandra, their median and 15.9th and 84.1th percentiles are shown. We have further derived $L_{2-10 \, \mathrm{keV}}$ and  $\alpha_{OX}$ from the this \textit{Chandra} flux posteriors. They are respectively marked in the left and right panels of Fig. \ref{fig:lxaox}.

\subsubsection{Fast outflows}
\label{sec:bal}
The discovery spectrum of J202040-621509 shows multi-trough, broad absorption lines (BALs). We first identify the following clear BALs in the quasar spectrum: NV, SiIV and CIV, which all show similar double-trough shapes \citep[e.g.][]{korista93}. In quasar spectra, BALs are imprinted by extreme outflows. Such outflows  are observed in about 12\% to 20\% of the quasar population \citep[e.g.][]{knigge08,scaringi09} and their velocity widths range can reach up to $\sim 30000 \, \mathrm{km\, s^{-1}}$ (so-called ultra-fast outflows, \citealt{rogerson16}). BALs are defined to have velocity widths $>2000 \, \mathrm{km\, s^{-1}}$ and reach maximal depths of at least 10\% below the quasar continuum level. This definition was introduced by  \citet{Weymann1991} to trace outflows while avoiding contamination by host and extra-galactic absorbing systems. As the NV trough is affected by the neutral IGM absorbtion shortwards of Lyman $\alpha$, we focus our analysis on the outflow velocity structure of CIV and SiIV. A measure of the trough widths, also called balnicity index (BI) was defined by  \citet{Weymann1991} as:

\begin{equation}
    \mathrm{BI} = \int_ {3000}^{25000} \left[ 1 - \frac{f(v)}{0.9} \right] C \, dv
\end{equation}.

Here $f(v)$ denotes the continuum-normalized flux, $C$ is a constant equal to 1 for velocities $>2000 \, \mathrm{km \, s^{-1}}$ with respect to the trough edge and 0 elsewhere. The measurement of the continuum is challenging due to the numerous BALs. We have approximated the continuum by using the principal component decomposition of $z\sim 3$ SDSS quasar continua presented by \citet{paris11}. The spectrum is strongly attenuated toward longer wavelengths, which is not accounted for by the PCA model. We fitted an additional power-law in the restframe spectral regions 1300-1350 $\AA$ and 1400-1600 $\AA$ to the spectrum normalized by the PCA-reprojected model. Finally, the spectrum of the source was normalised by the combined PCA and power-law models. We account for the flux uncertainties by bootstrapping from the spectrum RMS. From the normalized spectrum we measure $BI_\mathrm{CIV}$ = $3394 \pm  197 \, \mathrm{km\, s^{-1}}$ and $BI_\mathrm{SiIV}$ = $2935 \pm  110 \, \mathrm{km\, s^{-1}}$. This BAL strength confirms the classification of J202040-621509 as BAL quasar.

The large BALs are detached with respect to the quasar redshift, i.e. the troughs of all species are shifted shortwards of the expected corresponding quasar emission lines. We measure detached and terminal velocities from the start and the end of the absorption troughs. From the SiIV BAL we measure outflow velocities $v_\mathrm{SiIV}=0.002c-0.020c$ and  $v_\mathrm{CIV}=0.003c-0.025c$. We note that the optical spectrum and the X-ray data were not obtained simultaneously. The LDSS3 observations took place closer to the eRASS1 observation ($\sim 1$ year in the observed frame) than the \textit{Chandra} ACIS-S pointing ($\sim 2$ years), i.e. closer to the high-luminosity state discussed in Section \ref{sec:xvar}.

\subsection{Radio properties}
\label{sec:radiobal}

J202040$-$621509 is detected at radio frequencies, at 0.887 and 1.37 GHz in VAST and at 0.943 GHz in the EMU Pilot survey (\citealt{norris21}). 
We retrieved the corresponding radio images from CASDA and we performed a single Gaussian fit using the task IMFIT within the CASA software to quantify the flux densities (see Table \ref{J2020radio_fluxes}). 
The images are shown in Fig. \ref{radio_images} (top two rows) in Appendix \ref{append:radio}.
Several VAST images are available, corresponding to different observing dates. We extracted the flux density from each of them. 
By assuming a single power-law for the continuum radio emission, we obtained a radio spectral index ($\alpha_r$) of -0.07$\pm$0.15 (see Fig. \ref{radiospecs}) that makes J202040$-$621509 FSRQ, similar to J020916$-$562650.
The VAST measurements are consistent with each other, indicating that the source does not appear to be variable on time scales of a few months (observed frame).
Then we computed the values of radio-loudness (R), by following the same procedure as for J020916$-$562650.
We obtained a mean R of 10$\pm$4 which places the source at the boundary between objects classified as radio-quiet or radio-loud (\citealt{urry95}).
This could indicate the presence of a relativistic jet (e.g., \citealt{sbarrato21}).  

\subsection{Obscuring fast accretion disc winds at high redshift and other variability scenarios}

AGN shows substantial variability in flux and spectral shape at all frequencies (for a review, see \citealt{lira21}). This variability arises from accretion physics \citep[e.g.][]{nandra97}, broad line region kinematics \citep[e.g.][]{cherepashchuk73,boksenberg78,blandford82,peterson93} and jets \citep[e.g.]{albert07}.  In X-rays, AGN has been observed to be variable on timescales of hours to days \citep[e.g.][]{winkler75,marshall81,mchardy04,ponti12}. Little redshift evolution in the X-ray variabity of AGN, as traced by the variance-frequency diagram, has been observed from $z=0$ to $z=3$ \citep{paolillo23}. Interpreted as reverberation effect, i.e. reprocessing of accretion disk photons in the electron corona \citep{haardt91}, rapid X-ray variability can be used to study the structure of the innermost region of the AGN at a couple of gravitational radii away from the black hole \citep [e.g.][]{mchardy13}. Coronal variability usually shows maximum amplitudes of a factor $\sim 2$ \citep[e.g.,][]{gibson12}. However, more extreme variability, unrelated to coronal fluctuations, has been observed on longer timescales \citep[e.g.][]{timlin20}. For instance, \citep{lamassa15} reported the first discovery of a changing look quasar in which the AGN flux dropped by a factor of 6  over timescales of years while not showing absorption signatures. 
As presented in Section \ref{sec:bal}, the UV spectrum of J202040-621509, in contrast, is that of a typical BAL quasar.
The X-ray variability in quasars containing BALs in their restframe UV spectra is thought to arise from absorption effects. More precisely, variations in the column density of intervening, obscuring clouds are expected to arise from large-scale outflows \citep[e.g.][]{brandt00, green01, gallagher02,gallagher06,gibson09,saez12}. An often invoked physical model for the observed extreme outflow velocities is that of line-driven winds launched from the accretion disk, as high-accretion rates drive the UV radiation pressure of the disk up. High-density regions, located at the inner edge of this flow can shield the coronal X-ray emission 
 through photo-electric absorption (NH$>10^{22} \, \rm cm^{-2}$ \citealt{murray95}).

A possible scenario is that the observed X-ray dimming of J202040-621509 over one order of magnitude in X-ray flux may partly be explained by such obscuring winds. While too few counts have been collected in the \textit{Chandra} observation to extract and fit an X-ray spectrum, we only collected counts at energies $>6.7 \, \rm keV$ in the quasar rest frame. This hard photometry can, for example, be indicative of heavy X-ray absorption, affecting lower-energetic photons. 

Extreme X-ray variability in high-redshift quasars has been reported previously (e.g. at $z=5.41$, \citealt{shemmer05}).
Within the first gigayear, only two objects were known to display such a behaviour \citep{nanni18,vito22}. \citet{nanni18}
reported a hardening of the spectrum and a flux decrease of factor 2.5 in the quasar SDSS J1030+0524 ($z=6.31$) between two XMM-\textit{Newton} observations spaced by 14 years. They proposed intrinsic variation of accretion rate as a potential driver of the observed variability. \citet{vito22} presented the surprising non-detection in a 100ks XMM-\textit{Newton} observations of the quasar CFHQS J164121+375520 ($z = 6.047$), which had been detected 115 quasar-restframe days before with Chandra, implying a flux-dimming of factor $>7$. The authors discuss intervening obscuring material on sub-pc scales, such as disk-shape variability in super-Eddington sources. \citet{vito22} also list a possible model by \citet{giustini19}, in which rapidly accreting black holes with $\dot{m} \geq 0.25$, generate the favourable conditions for their optically-thick, thin disk emission to dominate the output of a cooled corona. These disks are expected to launch strong line-driven winds, imprinting BAL features in the spectrum of the quasars. 

While the latter model yields a promising scenario for the phenomenology observed in J202040-621509, we cannot exclude any of the above-mentioned sources of variability as the counting statistics of our X-ray observations are poor and we are missing simultaneous UV photometry to simultaneously track the corona and disk emission. A deeper, simultaneous X-ray and UV monitoring campaign will be required to shed light on this seemingly extreme X-ray dimming event.
\section{Conclusions}
\label{sec:six}

In this work, we have presented a pilot survey designed to uncover new X-ray luminous quasars at $z>5.6$, combining eROSITA data with optical and IR photometry from DES, VHS and CatWISE2020. We have discovered 5 quasars out of 9 sources that were selected for spectroscopic confirmation with LDSS3-C. Two of these sources are securely detected in eRASS. 
These sources are also radio-detected. J020916-562650 is formally radio-loud, while J202040-621509 lies right at the threshold delimiting radio-loud from radio-quiet sources ($R=10$).  
The reionisation-era quasars uncovered by our pilot study shed light on powerful X-ray emission mechanisms in the early universe and give us access to demographic diagnostics for the $z>5.6$ quasar population.

J020916-562650 ($z=5.6$) shows X-ray and radio properties of a blazar, making it one of the most distant AGN of this type known to date \citep{belladitta20,tao23,caccianiga24}. In their independent analysis, \citet{ighina24} also come to the conclusion that this source is a blazar. Recently, the discovery of a blazar at $z=7$ was reported by \citet{banados24}, setting a new redshift frontier for this type of source. From this detection and using simple geometric arguments, we infer a space density of similarly luminous, jetted quasars of $6.89^{15.86}_{-5.70}\, \mathrm{Gpc^{-3}}$. This value is consistent with the space density of all quasars, jetted or not, as measured by the most recent quasar luminosity functions at this redshift and brightness. Our measurement implies two potential scenarios: either all quasars discovered so far at $z\sim 6$ and M1450<-25 are jetted (and most of these jets have not been detected yet) or a substantial fraction of jetted quasars in this redshift-magnitude regime have eluded quasar searches so far, perhaps due to obscuration. Severe systematics that can affect our measurement of quasar space density are discussed in this paper.

J202040-621509,  ($z=5.7$) is an $R=10$ BAL quasar. Through 
 four consecutive eRASS observations and an additional \textit{Chandra} pointing, we provide substantial evidence for an extreme X-ray dimming event (a factor $\geq 13$ in X-ray flux). Potential dimming scenarios are nuclear obscuration or disk-dominated quasar SEDs in strongly outflowing quasars with rapidly accreting black holes. We conclude, however, that increased X-ray counting statistics and simultaneous UV monitoring will be required to draw clearer conclusions on this dimming event.

An eRASS hemisphere reionization quasar search will be based on this pilot study. With the advent of the next generation of wide-field optical and IR imaging surveys such as \textit{Euclid} Wide Survey \citep{euclid19,euclid24}, and surveys of the \textit{Rubin Observatory} \citep{rubin19} and the \textit{Nancy Grace Roman} Space Telescope, we will be able to explore new synergies with the eRASS, enabling the discovery of the most X-ray luminous quasars in the early universe.

\begin{acknowledgements}

     JW acknowledges support by the Deutsche Forschungsgemeinschaft (DFG, German Research Foundation) under Germany's Excellence Strategy - EXC-2094 - 390783311. JW is grateful to Dr. Eduardo Ba\~{n}ados for guidance on the topic of high-redshift quasar searches. He would also like to thank Dr. Peter Predehl and Dr. Vadim Burwitz for helping with the \textit{Chandra} observations presented in this project.
     
     \newline 
     T. Urrutia acknowledges funding from the ERC-AdG grant SPECMAP-CGM, GA 101020943.
    
     \newline
    
  This work is based on data from eROSITA, the soft X-ray instrument aboard SRG, a joint Russian-German science mission supported by the Russian Space Agency (Roskosmos), in the interests of the Russian Academy of Sciences represented by its Space Research Institute (IKI), and the Deutsches Zentrum für Luft- und Raumfahrt (DLR). The SRG spacecraft was built by Lavochkin Association (NPOL) and its subcontractors, and is operated by NPOL with support from the Max Planck Institute for Extraterrestrial Physics (MPE).

The development and construction of the eROSITA X-ray instrument was led by MPE, with contributions from the Dr. Karl Remeis Observatory Bamberg \& ECAP (FAU Erlangen-Nuernberg), the University of Hamburg Observatory, the Leibniz Institute for Astrophysics Potsdam (AIP), and the Institute for Astronomy and Astrophysics of the University of Tübingen, with the support of DLR and the Max Planck Society. The Argelander Institute for Astronomy of the University of Bonn and the Ludwig Maximilians Universität Munich also participated in the science preparation for eROSITA.

The eROSITA data shown here were processed using the eSASS/NRTA software system developed by the German eROSITA consortium.

     \newline

      The scientific results reported in this article are based to a significant degree on observations made by the \textit{Chandra} X-ray Observatory.

    \newline

his project used public archival data from the Dark Energy Survey (DES). Funding for the DES Projects has been provided by the U.S. Department of Energy, the U.S. National Science Foundation, the Ministry of Science and Education of Spain, the Science and Technology Facilities Council of the United Kingdom, the Higher Education Funding Council for England, the National Center for Supercomputing Applications at the University of Illinois at Urbana–Champaign, the Kavli Institute of Cosmological Physics at the University of Chicago, the Center for Cosmology and Astro-Particle Physics at the Ohio State University, the Mitchell Institute for Fundamental Physics and Astronomy at Texas A\&M University, Financiadora de Estudos e Projetos, Fundação Carlos Chagas Filho de Amparo à Pesquisa do Estado do Rio de Janeiro, Conselho Nacional de Desenvolvimento Científico e Tecnológico and the Ministério da Ciência, Tecnologia e Inovação, the Deutsche Forschungsgemeinschaft and the Collaborating Institutions in the Dark Energy Survey.

The Collaborating Institutions are Argonne National Laboratory, the University of California at Santa Cruz, the University of Cambridge, Centro de Investigaciones Enérgeticas, Medioambientales y Tecnol$\acute{o}$gicas–Madrid, the University of Chicago, University College London, the DES-Brazil Consortium, the University of Edinburgh, the Eidgenössische Technische Hochschule (ETH) Zürich, Fermi National Accelerator Laboratory, the University of Illinois at Urbana-Champaign, the Institut de Ciències de l’Espai (IEEC/CSIC), the Institut de Física d’Altes Energies, Lawrence Berkeley National Laboratory, the Ludwig-Maximilians Universität München and the associated Excellence Cluster Universe, the University of Michigan, the National Optical Astronomy Observatory, the University of Nottingham, The Ohio State University, the OzDES Membership Consortium, the University of Pennsylvania, the University of Portsmouth, SLAC National Accelerator Laboratory, Stanford University, the University of Sussex, and Texas A\&M University.

Based in part on observations at Cerro Tololo Inter-American Observatory, National Optical Astronomy Observatory, which is operated by the Association of Universities for Research in Astronomy (AURA) under a cooperative agreement with the National Science Foundation.

\newline
CatWISE uses data products from WISE, which is a joint project of the University of California, Los Angeles, and the Jet Propulsion Laboratory (JPL)/California Institute of Technology (Caltech), funded by the National Aeronautics and Space Administration (NASA), and from NEOWISE, which is a joint project of JPL/Caltech and the University of Arizona funded by NASA. CatWISE is led by JPL/Caltech, with funding from NASA's Astrophysics Data Analysis Program.
\newline 
Based on observations obtained as part of the VISTA Hemisphere Survey, ESO Progam, 179.A-2010 (PI: McMahon)
\newline
The Legacy Surveys consist of three individual and complementary projects: the Dark Energy Camera Legacy Survey (DECaLS; Proposal ID $\#$2014B-0404; PIs: David Schlegel and Arjun Dey), the Beijing-Arizona Sky Survey (BASS; NOAO Prop. ID \#2015A-0801; PIs: Zhou Xu and Xiaohui Fan), and the Mayall z-band Legacy Survey (MzLS; Prop. ID \#2016A-0453; PI: Arjun Dey). DECaLS, BASS and MzLS together include data obtained, respectively, at the Blanco telescope, Cerro Tololo Inter-American Observatory, NSF’s NOIRLab; the Bok telescope, Steward Observatory, University of Arizona; and the Mayall telescope, Kitt Peak National Observatory, NOIRLab. Pipeline processing and analyses of the data were supported by NOIRLab and the Lawrence Berkeley National Laboratory (LBNL). The Legacy Surveys project is honored to be permitted to conduct astronomical research on Iolkam Du’ag (Kitt Peak), a mountain with particular significance to the Tohono O’odham Nation.

NOIRLab is operated by the Association of Universities for Research in Astronomy (AURA) under a cooperative agreement with the National Science Foundation. LBNL is managed by the Regents of the University of California under contract to the U.S. Department of Energy.

This project used data obtained with the Dark Energy Camera (DECam), which was constructed by the Dark Energy Survey (DES) collaboration. Funding for the DES Projects has been provided by the U.S. Department of Energy, the U.S. National Science Foundation, the Ministry of Science and Education of Spain, the Science and Technology Facilities Council of the United Kingdom, the Higher Education Funding Council for England, the National Center for Supercomputing Applications at the University of Illinois at Urbana-Champaign, the Kavli Institute of Cosmological Physics at the University of Chicago, Center for Cosmology and Astro-Particle Physics at the Ohio State University, the Mitchell Institute for Fundamental Physics and Astronomy at Texas A\&M University, Financiadora de Estudos e Projetos, Fundacao Carlos Chagas Filho de Amparo, Financiadora de Estudos e Projetos, Fundacao Carlos Chagas Filho de Amparo a Pesquisa do Estado do Rio de Janeiro, Conselho Nacional de Desenvolvimento Cientifico e Tecnologico and the Ministerio da Ciencia, Tecnologia e Inovacao, the Deutsche Forschungsgemeinschaft and the Collaborating Institutions in the Dark Energy Survey. The Collaborating Institutions are Argonne National Laboratory, the University of California at Santa Cruz, the University of Cambridge, Centro de Investigaciones Energeticas, Medioambientales y Tecnologicas-Madrid, the University of Chicago, University College London, the DES-Brazil Consortium, the University of Edinburgh, the Eidgenossische Technische Hochschule (ETH) Zurich, Fermi National Accelerator Laboratory, the University of Illinois at Urbana-Champaign, the Institut de Ciencies de l’Espai (IEEC/CSIC), the Institut de Fisica d’Altes Energies, Lawrence Berkeley National Laboratory, the Ludwig Maximilians Universitat Munchen and the associated Excellence Cluster Universe, the University of Michigan, NSF’s NOIRLab, the University of Nottingham, the Ohio State University, the University of Pennsylvania, the University of Portsmouth, SLAC National Accelerator Laboratory, Stanford University, the University of Sussex, and Texas A\&M University.

BASS is a key project of the Telescope Access Program (TAP), which has been funded by the National Astronomical Observatories of China, the Chinese Academy of Sciences (the Strategic Priority Research Program "The Emergence of Cosmological Structures" Grant \# XDB09000000), and the Special Fund for Astronomy from the Ministry of Finance. The BASS is also supported by the External Cooperation Program of Chinese Academy of Sciences (Grant \# 114A11KYSB20160057), and Chinese National Natural Science Foundation (Grant \# 12120101003, \# 11433005).

The Legacy Survey team makes use of data products from the Near-Earth Object Wide-field Infrared Survey Explorer (NEOWISE), which is a project of the Jet Propulsion Laboratory/California Institute of Technology. NEOWISE is funded by the National Aeronautics and Space Administration.

The Legacy Surveys imaging of the DESI footprint is supported by the Director, Office of Science, Office of High Energy Physics of the U.S. Department of Energy under Contract No. DE-AC02-05CH1123, by the National Energy Research Scientific Computing Center, a DOE Office of Science User Facility under the same contract; and by the U.S. National Science Foundation, Division of Astronomical Sciences under Contract No. AST-0950945 to NOAO.

\newline 
 
    This research made use of Astropy,\footnote{http://www.astropy.org} a community-developed core Python package for Astronomy \citep{astropy13, astropy18}. 
    \newline
      This research has made use of software provided by the \textit{Chandra} X-ray Center (CXC) in the application packages CIAO and Sherpa.
      \newline

      This research made use of Astropy,3
a community-developed core Python
package for Astronomy \citep{astropy13,astropy18}.

\newline
This research made use of Photutils, an Astropy package for
detection and photometry of astronomical sources \citep{photutils}.
This research has made use of "Aladin sky atlas" developed at CDS, Strasbourg Observatory, France
\newline
    \newline
    In addition this research made use of BXA (https://johannesbuchner.github.io/BXA/, \citealt{buchner14}), the \texttt{corner} package \citep{corner}, Pymoc, Aladin \citep{aladin}, topcat \citep{topcat}, lmfit \citep{lmfit}, seaborn, cmocean, matplotlib

\end{acknowledgements}%
--------------------------------------------------------------------

\bibliographystyle{aa}
\bibliography{bibliography}

\begin{appendix}

\section{Example DES images and forced photometry: Discarded quasar candidates}
\label{append:images}

\begin{figure*}
    
    \centering
    \includegraphics[width=1.99\columnwidth]{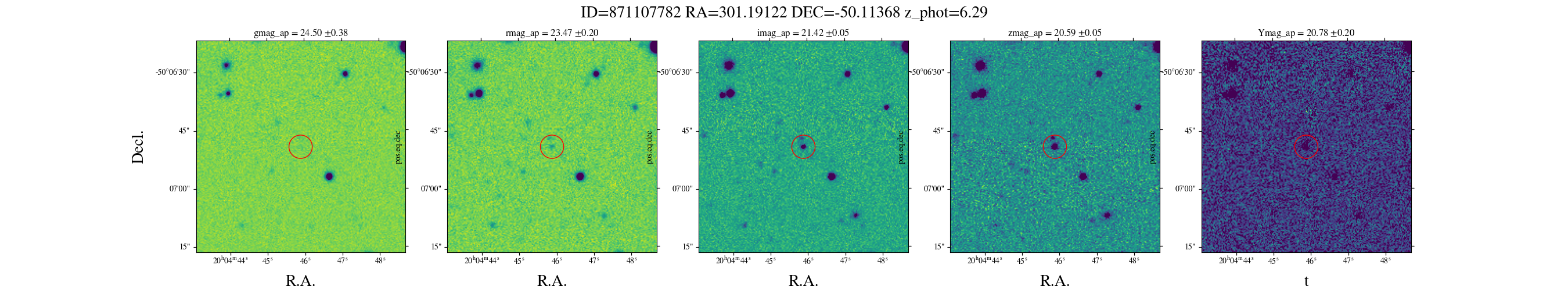}
    \includegraphics[width=1.99\columnwidth]{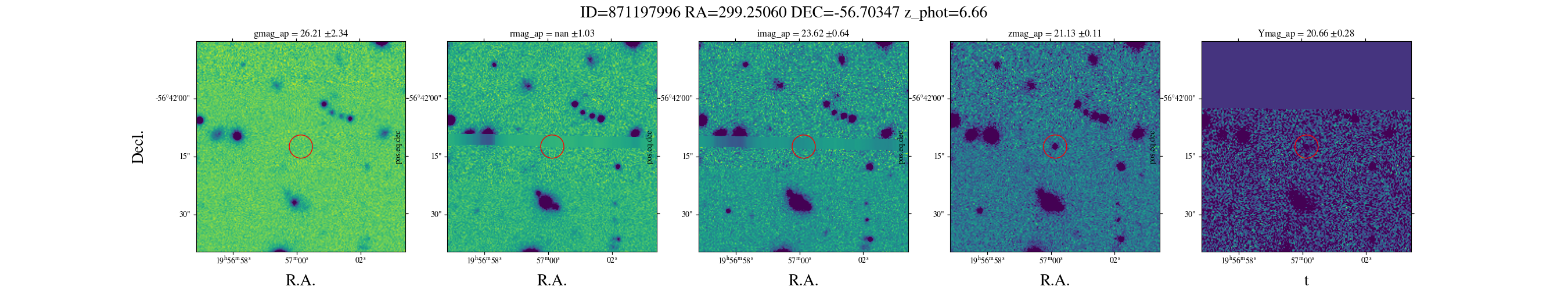}
    \includegraphics[width=1.99\columnwidth]{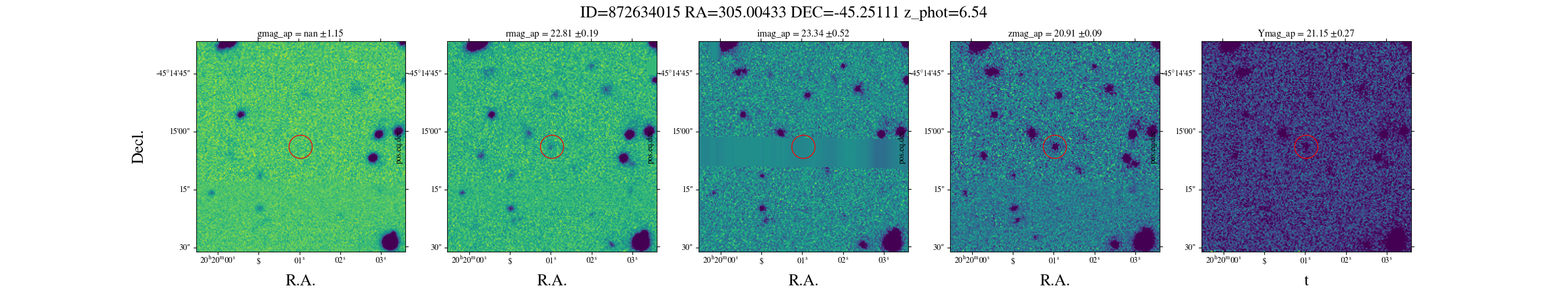}

    \caption{Discarded candidates after visual inspection and forced photometry.}
    \label{fig:contam_test}
\end{figure*}

In Fig. \ref{fig:contam_test}, we present example DES cutouts of candidates that were selected and candidates that were rejected after visual inspection and forced photometry. While the first example is a close pair, the second and the third are affected by artifacts such as bleed-trails and missing images. 

\section{Contaminant spectra and coordinates}
\label{append:contaminants}

The coordinates of the contaminants observed with LDSS3 are listed in Table \ref{tab:contam}.

\begin{table}[h]
\caption{Coordinates of the contaminants that were spectroscopically confirmed during the calibration of the quasar discovery pipeline.}
\begin{tabular}{@{}lll@{}}
\toprule
R.A.          & Decl.          & Cont. type \\ \midrule
05:38:50.9316 & -58:37:03.9792 & Cool dwarf \\
04:24:46.236  & -41:32:15.4788 & Cool dwarf \\
05:29:35.5414 & -19:27:34.362  & Cool dwarf \\
20:54:02.033  & -52:52:13.004  & Cool dwarf \\ \bottomrule
\end{tabular}
\tablefoot{ The spectral type of these sources is inferred by-eye.}
\label{tab:contam}

\end{table}

\section{Radio flux densities and radio images}
\label{append:radio}

\begin{figure*}
    \centering
    \includegraphics[width=1.99\columnwidth]{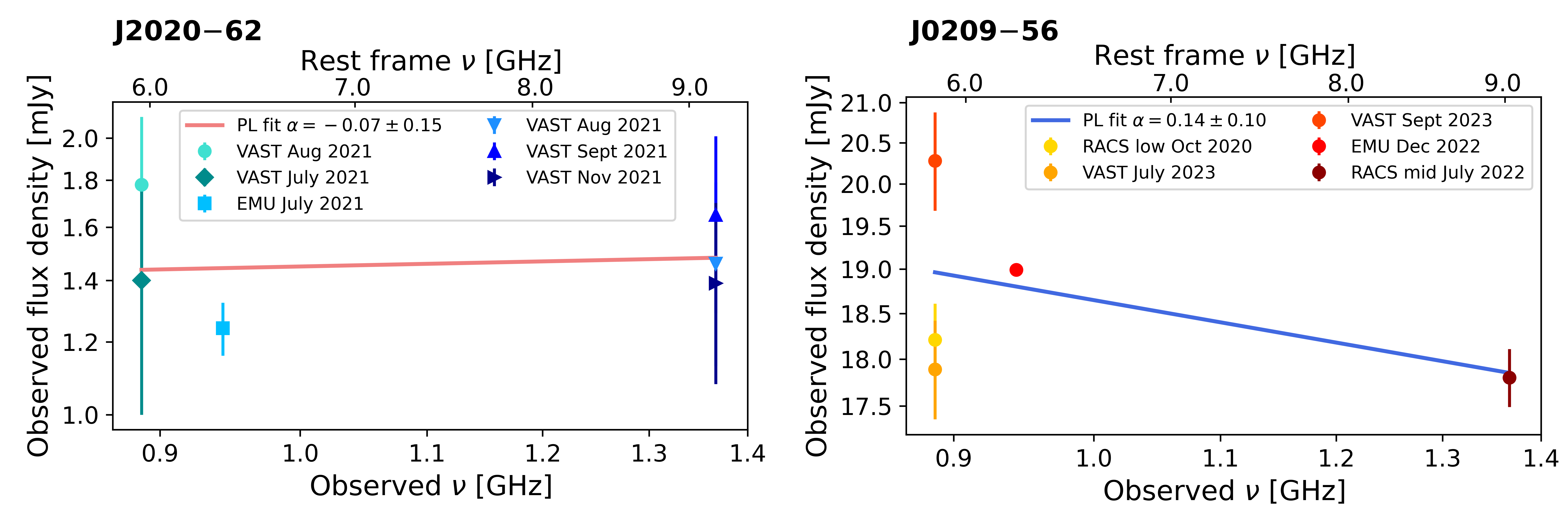}
   \caption{\small Observed radio spectral energy distribution of J202040$-$621509 (\textit{left} and J020916$-$562650 (\textit{right} from 0.887 to 1.37 GHz. The estimated flux densities are reported in the legend, together with the date of observation and the value of the best fit spectral index. The radio data of J020916$-$562650 at 0.887 GHz suggest a hint of variability. Simultaneous radio observations are crucial to reveal the true radio spectral energy distribution.     }
    \label{radiospecs}
\end{figure*}

Two of the newly discovered quasars are detected at radio frequencies in several surveys carried out with the Australian SKA Pathfinder (ASKAP, see the text in Section \ref{sec:five} for all the details). Here we list the available flux densities and radio images for these two objects (J2020$-$6215 and J0209$-$5626). 

\begin{small}
 \begin{table}[!ht]
 \caption{Summary of the archival radio observations of J202040$-$621509.}
 \label{J2020radio_fluxes}
 \centering
 \begin{tabular}{p{0.8cm}p{1.7cm}p{0.9cm}p{1.3cm}p{1.7cm}}
 \hline\hline
 Obs. Freq. & S$_{\nu}$ & Survey & Resolution & Obs. date \\
     (GHz)     & (mJy)       &             & (arcsec)      &                  \\
 (1) & (2) & (3) & (4) & (5) \\
 \hline
0.8875    & 1.78$\pm$0.33        &   VAST  & 20 & 2021-08-24 \\
0.8875    & 1.40$\pm$0.40        &   VAST  & 20 & 2021-07-24 \\
0.9435    & 1.242$\pm$0.082    &   EMU   & 10 &  2019-07-17 \\
1.367      & 1.46$\pm$0.33        &   VAST  & 10 &  2021-08-01 \\
1.367      & 1.65$\pm$0.36        &   VAST  & 10 &  2021-09-22 \\
1.367      & 1.39$\pm$0.31        &   VAST  & 10 &  2021-11-19  \\
\hline
 \end{tabular}
 \tablefoot{ Col(1): Observed frequency in GHz; Col(2): integrated flux density in mJy; Col(3): reference survey; Col(4): angular resolution in arcsec; Col(5): Date of the observation.} \end{table}
\end{small}

\begin{small}
 \begin{table}[!ht]
 \caption{Summary of the archival radio observations of J020916$-$562650.}
 \label{J0209radio_fluxes}
 \centering
 \begin{tabular}{p{0.8cm}p{1.7cm}p{0.9cm}p{1.3cm}p{1.7cm}}
 \hline\hline
 Obs. Freq. & S$_{\nu}$ & Survey & Resolution & Obs. date \\
     (GHz)     & (mJy)       &             & (arcsec)      &                  \\
 (1) & (2) & (3) & (4) & (5) \\
 \hline
  0.8875    &  18.21$\pm$0.44 &  RACS  & 25.0  &  2020-10-17 \\
  0.8875    &  17.89$\pm$0.56 &  VAST           & 13.0  &  2023-08-07 \\
  0.8875    &  20.28$\pm$0.63 &  VAST           & 13.0  &  2023-09-04 \\
  0.9435    &  18.99$\pm$0.20 &  EMU            & 15.0  &  2022-12-27 \\
  1.367      &  17.8$\pm$ 0.36 &  RACS & 10.0  &  2022-07-30 \\ 
\hline
 \end{tabular}
 \tablefoot{ Col(1): Observed frequency in GHz; Col(2): integrated flux density in mJy; Col(3): reference survey; Col(4): angular resolution in arcsec; Col(5): Date of the observation.} 
\end{table}
\end{small}

\begin{figure*}[!ht]
\centering
   {\includegraphics[width=5.0cm]{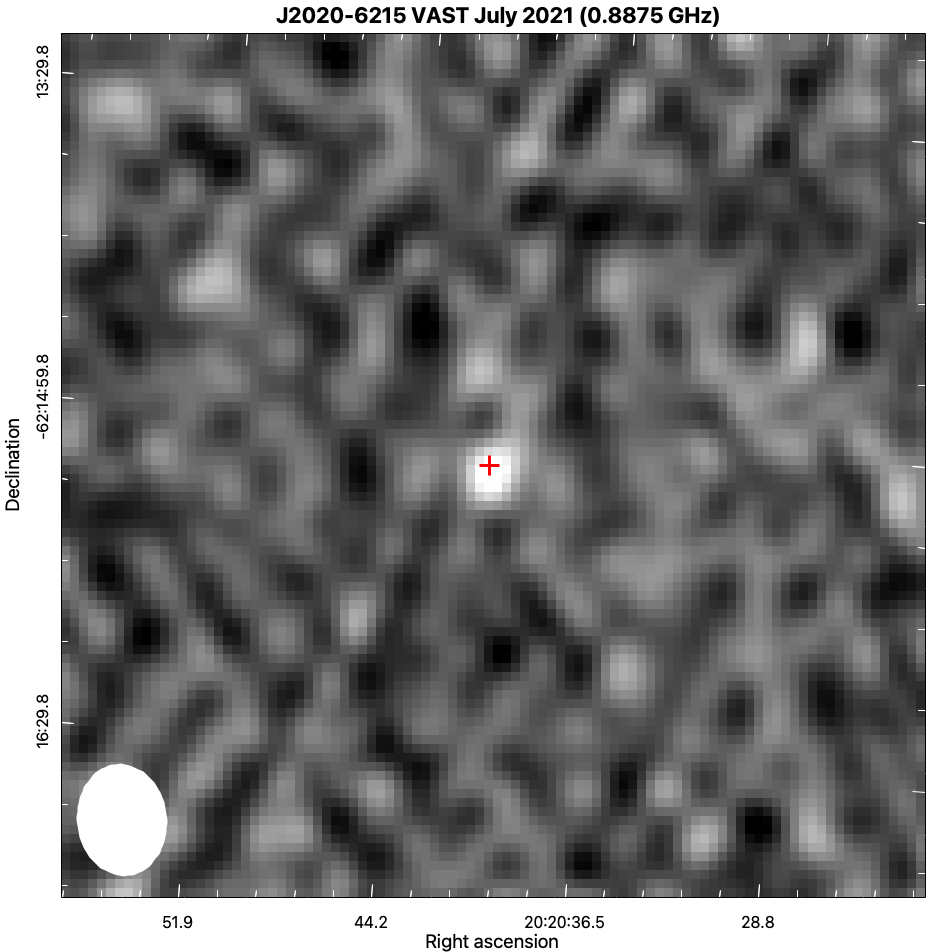}\hspace{0.5cm}
   \includegraphics[width=5.0cm]{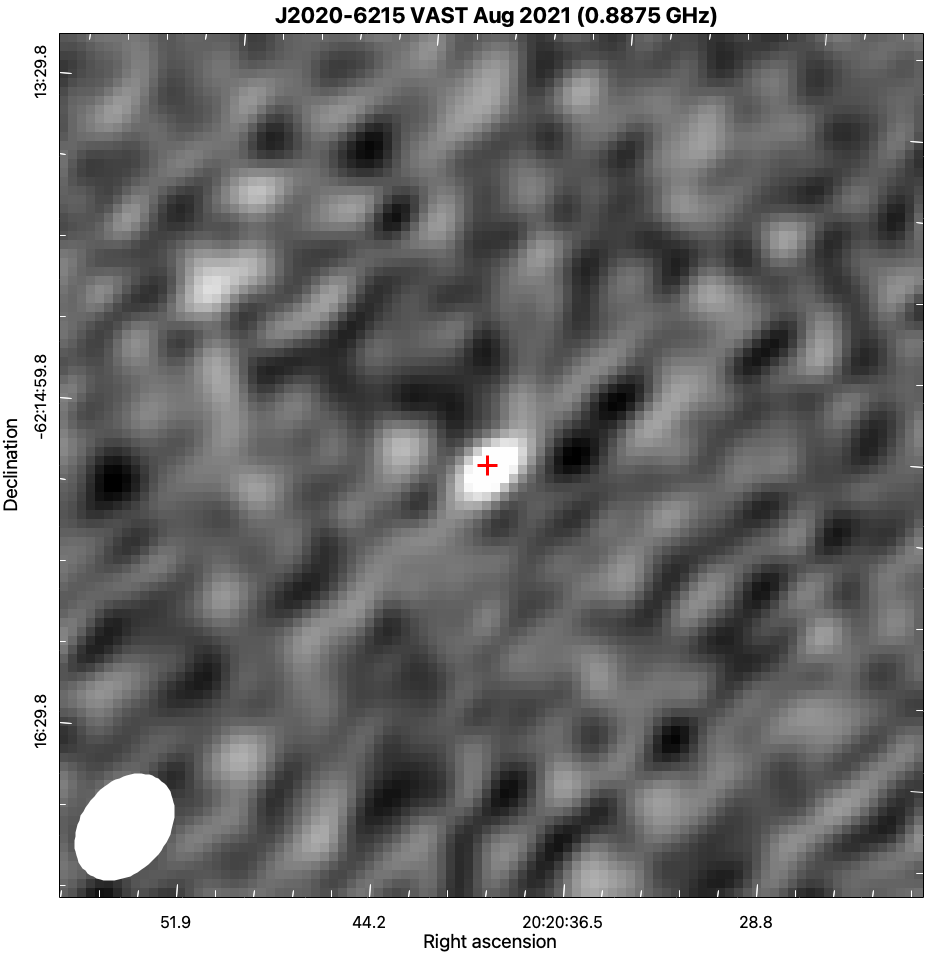}\hspace{0.5cm}
   \includegraphics[width=5.0cm]{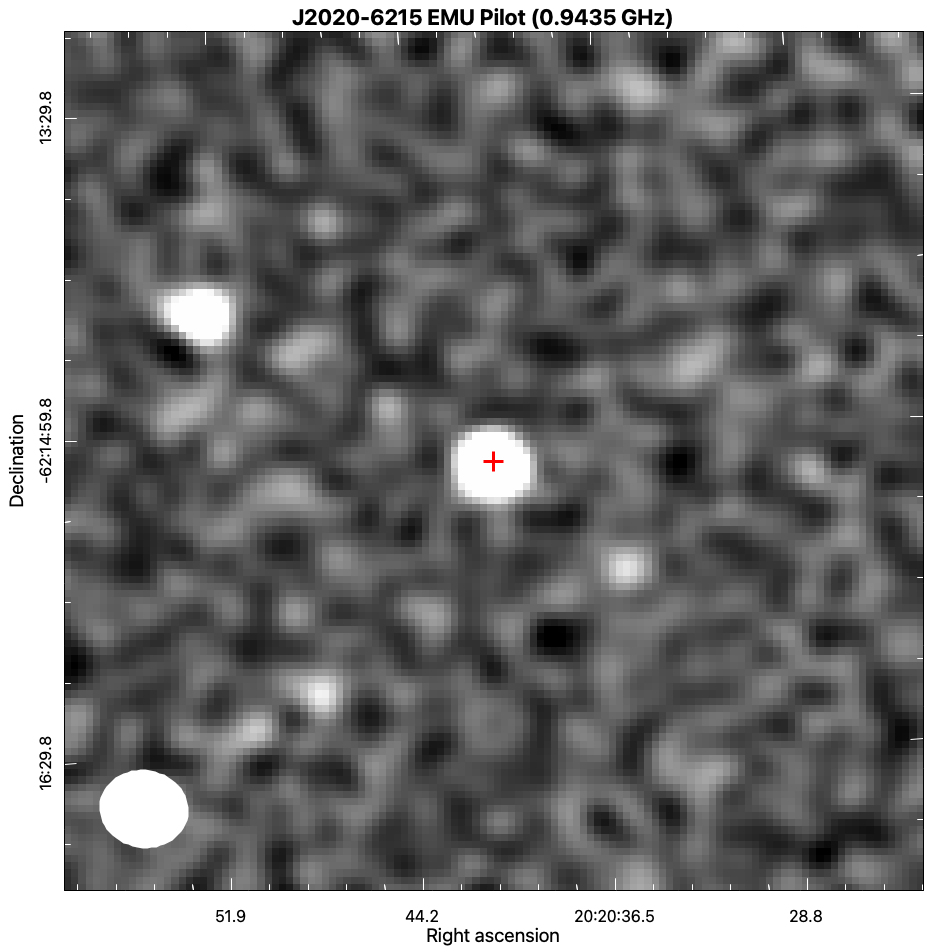}\hspace{0.5cm}
   \includegraphics[width=5.0cm]{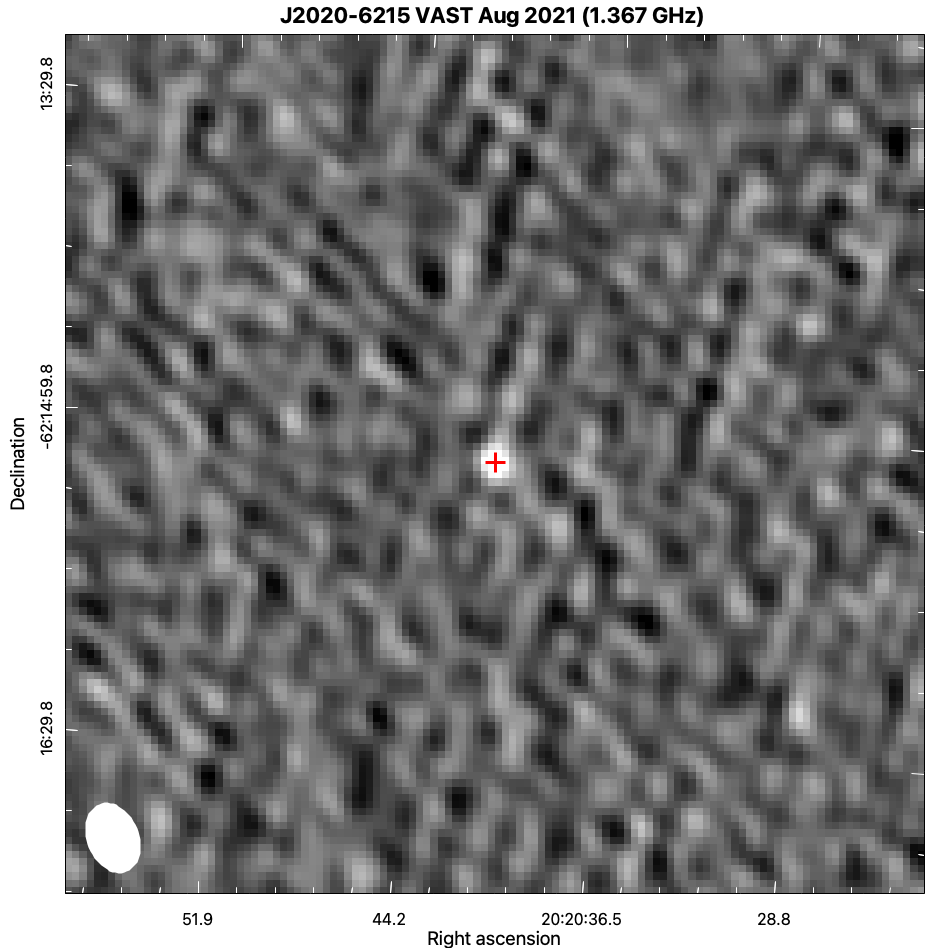}\hspace{0.5cm}
   \includegraphics[width=5.0cm]{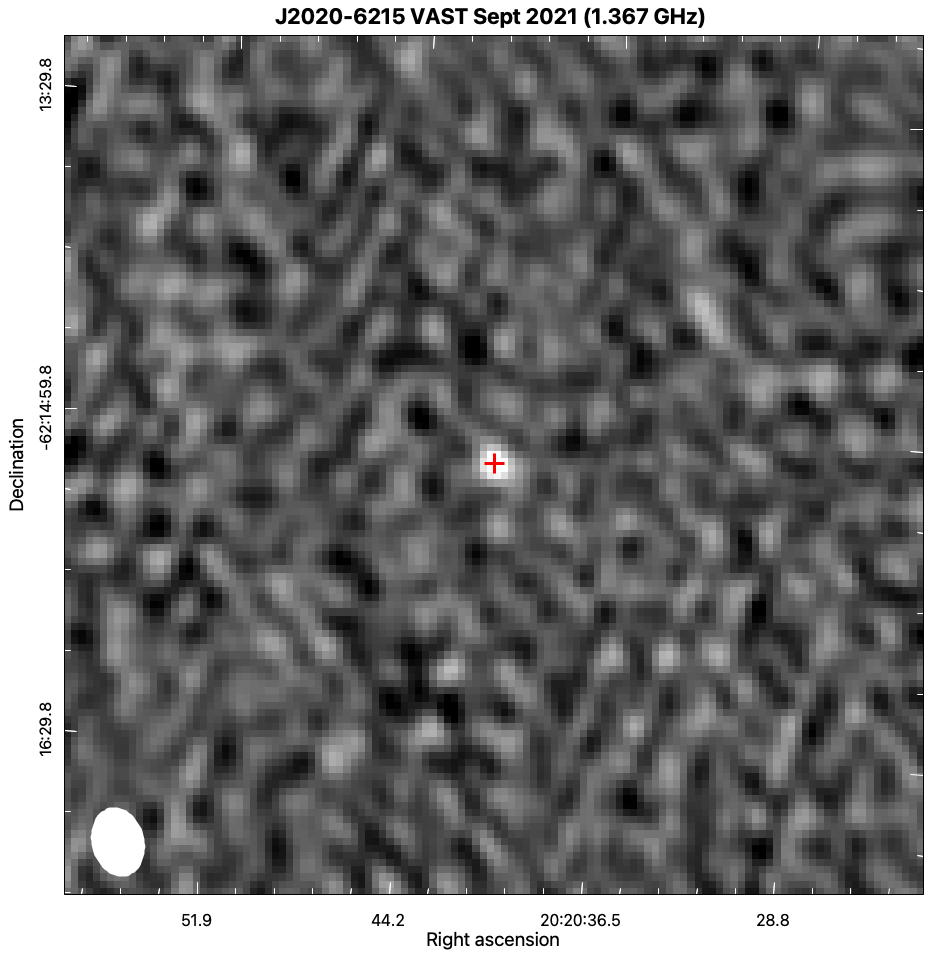}\hspace{0.5cm}
   \includegraphics[width=5.0cm]{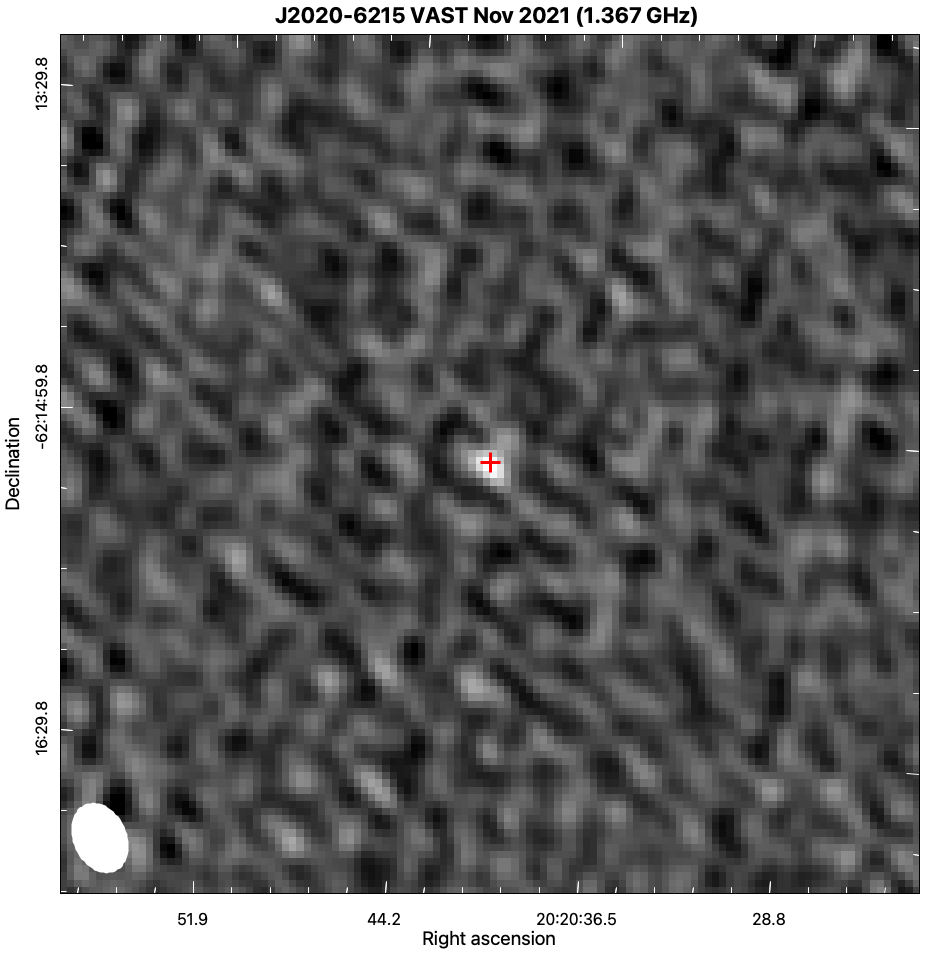}\hspace{1.5cm}
   \includegraphics[width=5.0cm]{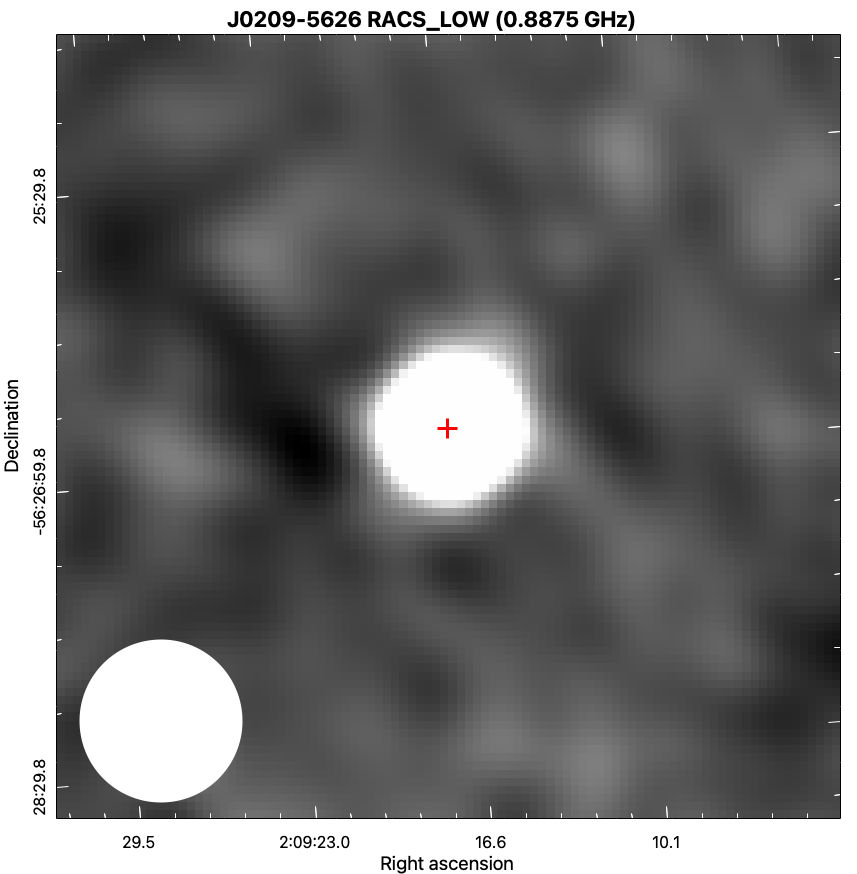}\hspace{0.5cm}
   \includegraphics[width=5.0cm]{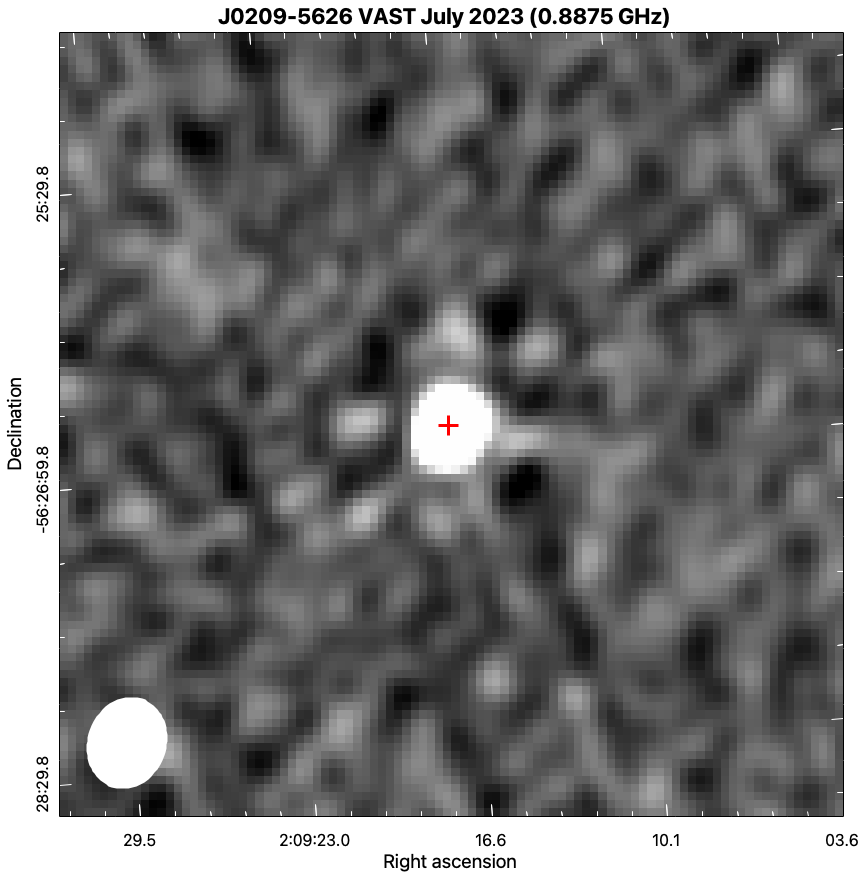}\hspace{0.5cm}
   \includegraphics[width=5.0cm]{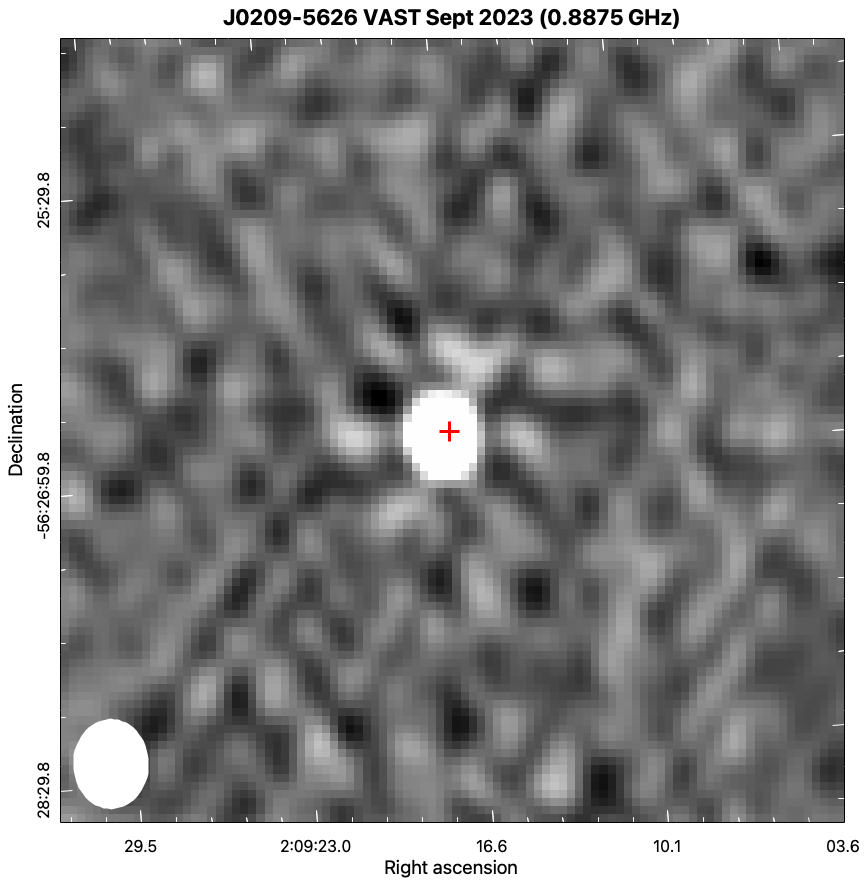}\hspace{0.5cm}
   \includegraphics[width=5.0cm]{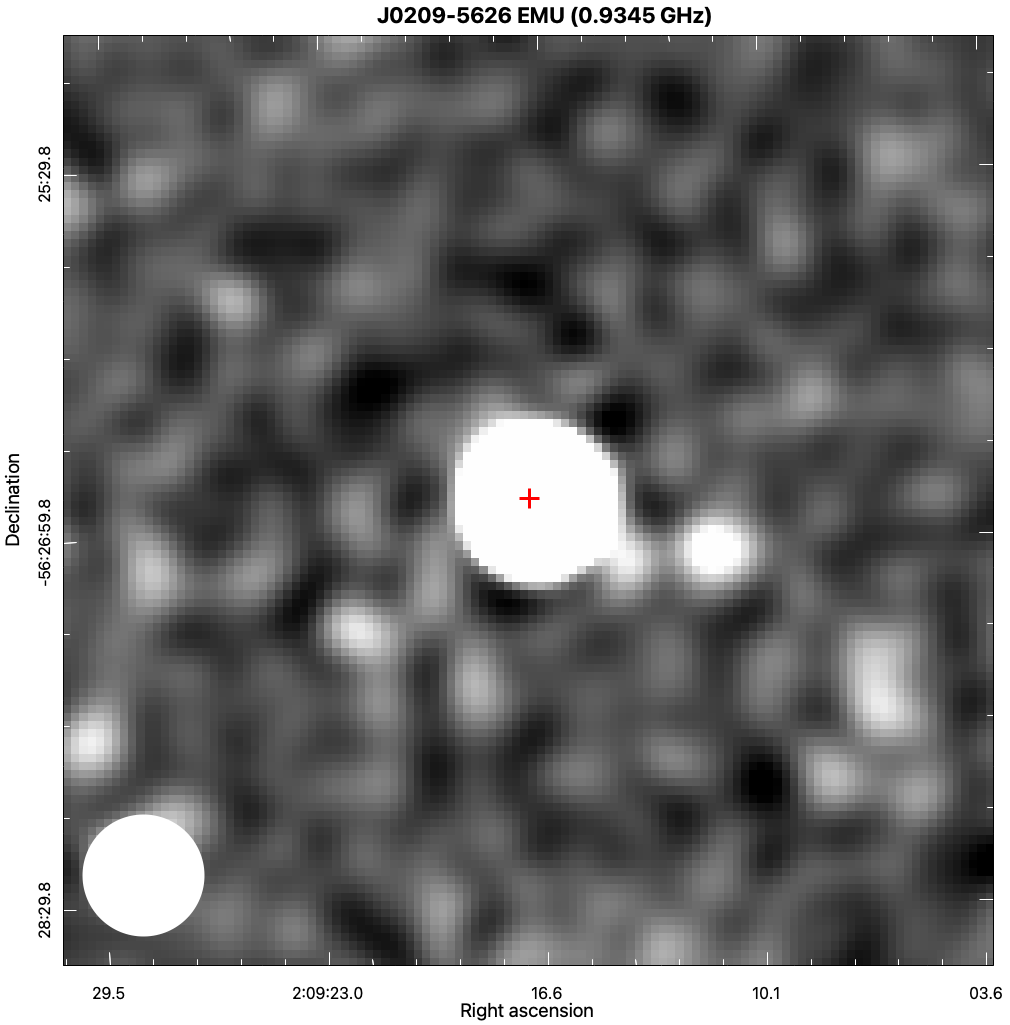}\hspace{0.5cm}
   \includegraphics[width=5.0cm]{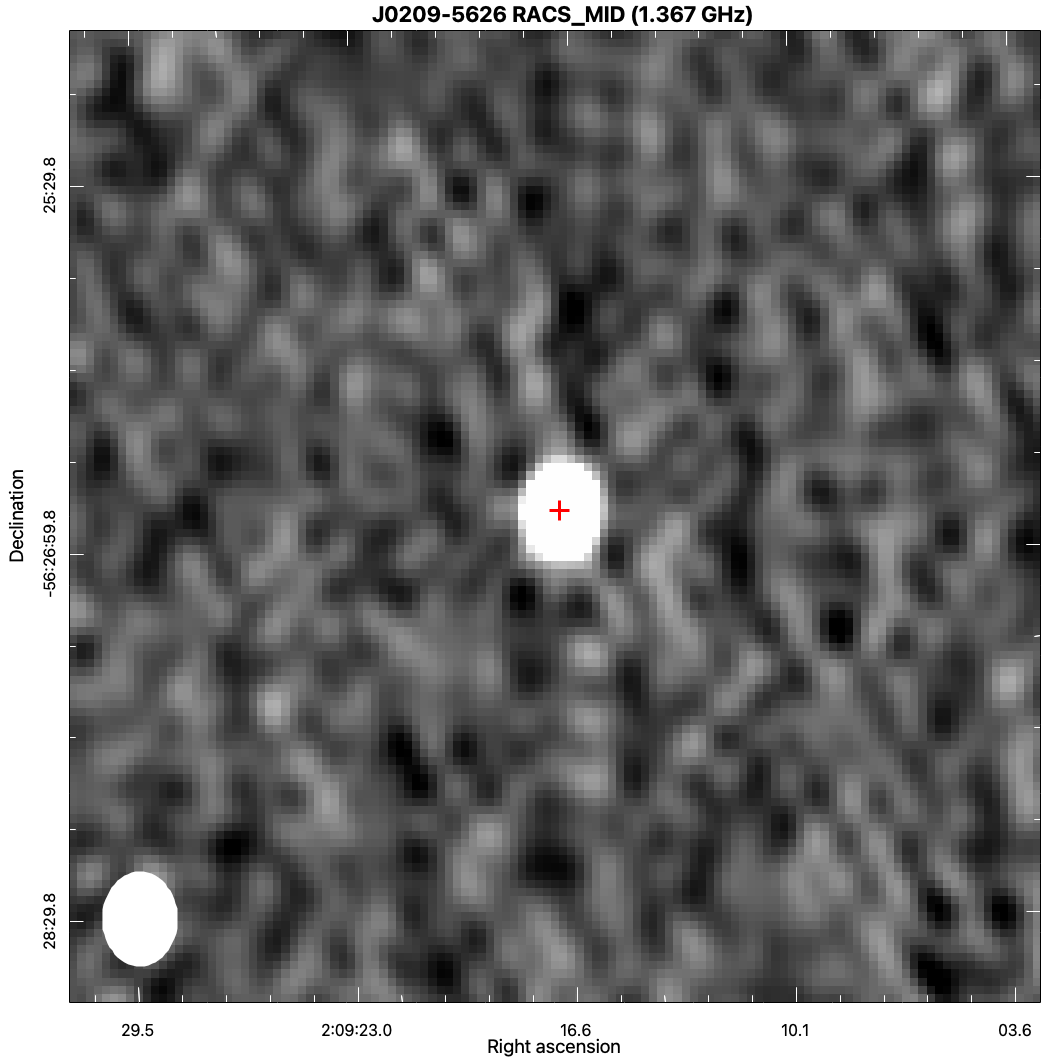}}
\caption{\small 4$’$$\times$4$’$ cutouts radio images of J2020$-$6215 (first and second row) and J0209$-$5626 (third and fourth row), ordered in increasing frequency. Both sources are clearly detected in all the images. The red cross represents the position of the optical source, and the synthesised beam of each observation is shown in the bottom left-hand corner.} 
\label{radio_images}
\end{figure*}
\end{appendix}

\end{document}